\documentclass[submission,copyright]{eptcs}
\providecommand{\event}{SAIRP 2013} 

\title{Notions of Monad Strength}
\author{Philip Mulry
\institute{Department of Computer Science\\
Colgate University\\
Hamilton, New York, 13346, USA}
\email{pmulry@colgate.edu}
}
\def\titlerunning{Monad Strength}
\def\authorrunning{P. S. Mulry}

\begin{document}
\maketitle

\makeatletter

\def\diagram{\leftwidth=\z@ \rightwidth=\z@ \topheight=\z@
\botheight=\z@ \setbox\@picbox\hbox\bgroup}

\def\enddiagram{\egroup\wd\@picbox\rightwidth\unitlength
\ht\@picbox\topheight\unitlength \dp\@picbox\botheight\unitlength
\hskip\leftwidth\unitlength\box\@picbox}

\def\bfig{\begin{diagram}}
\def\efig{\end{diagram}}
\newcount\wideness \newcount\leftwidth \newcount\rightwidth
\newcount\highness \newcount\topheight \newcount\botheight

\def\ratchet#1#2{\ifnum#1<#2 \global #1=#2 \fi}

\def\putbox(#1,#2)#3{%
\horsize{\wideness}{#3} \divide\wideness by 2
{\advance\wideness by #1 \ratchet{\rightwidth}{\wideness}}
{\advance\wideness by -#1 \ratchet{\leftwidth}{\wideness}}
\vertsize{\highness}{#3} \divide\highness by 2
{\advance\highness by #2 \ratchet{\topheight}{\highness}}
{\advance\highness by -#2 \ratchet{\botheight}{\highness}}
\put(#1,#2){\makebox(0,0){$#3$}}}

\def\putlbox(#1,#2)#3{%
\horsize{\wideness}{#3}
{\advance\wideness by #1 \ratchet{\rightwidth}{\wideness}}
{\ratchet{\leftwidth}{-#1}}
\vertsize{\highness}{#3} \divide\highness by 2
{\advance\highness by #2 \ratchet{\topheight}{\highness}}
{\advance\highness by -#2 \ratchet{\botheight}{\highness}}
\put(#1,#2){\makebox(0,0)[l]{$#3$}}}

\def\putrbox(#1,#2)#3{%
\horsize{\wideness}{#3}
{\ratchet{\rightwidth}{#1}}
{\advance\wideness by -#1 \ratchet{\leftwidth}{\wideness}}
\vertsize{\highness}{#3} \divide\highness by 2
{\advance\highness by #2 \ratchet{\topheight}{\highness}}
{\advance\highness by -#2 \ratchet{\botheight}{\highness}}
\put(#1,#2){\makebox(0,0)[r]{$#3$}}}

\def\adjust[#1]{} 

\newcount \coefa
\newcount \coefb
\newcount \coefc
\newcount\tempcounta
\newcount\tempcountb
\newcount\tempcountc
\newcount\tempcountd
\newcount\xext
\newcount\yext
\newcount\xoff
\newcount\yoff
\newcount\gap%
\newcount\arrowtypea
\newcount\arrowtypeb
\newcount\arrowtypec
\newcount\arrowtyped
\newcount\arrowtypee
\newcount\height
\newcount\width
\newcount\xpos
\newcount\ypos
\newcount\run
\newcount\rise
\newcount\arrowlength
\newcount\halflength
\newcount\arrowtype
\newdimen\tempdimen
\newdimen\xlen
\newdimen\ylen
\newsavebox{\tempboxa}%
\newsavebox{\tempboxb}%
\newsavebox{\tempboxc}%

\newdimen\w@dth

\def\setw@dth#1#2{\setbox\z@\hbox{$#1$}\w@dth=\wd\z@
\setbox\@ne\hbox{$#2$}\ifnum\w@dth<\wd\@ne \w@dth=\wd\@ne \fi
\advance\w@dth by 1.2em}


\def\t@^#1_#2{\def\n@one{#1}\def\n@two{#2}\mathrel{\setw@dth{#1}{#2}
\mathop{\hbox to \w@dth{\rightarrowfill}}\limits
\ifx\n@one\empty\else ^{\box\z@}\fi
\ifx\n@two\empty\else _{\box\@ne}\fi}}
\def\t@@^#1{\@ifnextchar_ {\t@^{#1}}{\t@^{#1}_{}}}
\def\to{\@ifnextchar^ {\t@@}{\t@@^{}}}

\def\t@left^#1_#2{\def\n@one{#1}\def\n@two{#2}\mathrel{\setw@dth{#1}{#2}
\mathop{\hbox to \w@dth{\leftarrowfill}}\limits
\ifx\n@one\empty\else ^{\box\z@}\fi
\ifx\n@two\empty\else _{\box\@ne}\fi}}
\def\t@@left^#1{\@ifnextchar_ {\t@left^{#1}}{\t@left^{#1}_{}}}
\def\toleft{\@ifnextchar^ {\t@@left}{\t@@left^{}}}

\def\two@^#1_#2{\def\n@one{#1}\def\n@two{#2}\mathrel{\setw@dth{#1}{#2}
\mathop{\vcenter{\hbox to \w@dth{\rightarrowfill}\kern-1.7ex
                 \hbox to \w@dth{\rightarrowfill}}%
       }\limits
\ifx\n@one\empty\else ^{\box\z@}\fi
\ifx\n@two\empty\else _{\box\@ne}\fi}}
\def\tw@@^#1{\@ifnextchar_ {\two@^{#1}}{\two@^{#1}_{}}}
\def\two{\@ifnextchar^ {\tw@@}{\tw@@^{}}}

\def\tofr@^#1_#2{\def\n@one{#1}\def\n@two{#2}\mathrel{\setw@dth{#1}{#2}
\mathop{\vcenter{\hbox to \w@dth{\rightarrowfill}\kern-1.7ex
                 \hbox to \w@dth{\leftarrowfill}}%
       }\limits
\ifx\n@one\empty\else ^{\box\z@}\fi
\ifx\n@two\empty\else _{\box\@ne}\fi}}
\def\t@fr@^#1{\@ifnextchar_ {\tofr@^{#1}}{\tofr@^{#1}_{}}}
\def\tofro{\@ifnextchar^ {\t@fr@}{\t@fr@^{}}}

\def\epi{\mathop{\mathchar"221\mkern -12mu\mathchar"221}\limits}
\def\leftepi{\mathop{\mathchar"220\mkern -12mu\mathchar"220}\limits}
\def\mon{\mathop{\m@th\hbox to
      14.6\P@{\lasyb\char'51\hskip-2.1\P@$\arrext$\hss
$\mathord\rightarrow$}}\limits} 
\def\leftmono{\mathrel{\m@th\hbox to
14.6\P@{$\mathord\leftarrow$\hss$\arrext$\hskip-2.1\P@\lasyb\char'50%
}}\limits} 
\mathchardef\arrext="0200       

\setlength{\unitlength}{.01em}%
\def\settypes(#1,#2,#3){\arrowtypea#1 \arrowtypeb#2 \arrowtypec#3}
\def\settoheight#1#2{\setbox\@tempboxa\hbox{#2}#1\ht\@tempboxa\relax}%
\def\settodepth#1#2{\setbox\@tempboxa\hbox{#2}#1\dp\@tempboxa\relax}%
\def\settokens[#1`#2`#3`#4]{%
     \def\tokena{#1}\def\tokenb{#2}\def\tokenc{#3}\def\tokend{#4}}
\def\setsqparms[#1`#2`#3`#4;#5`#6]{%
\arrowtypea #1
\arrowtypeb #2
\arrowtypec #3
\arrowtyped #4
\width #5
\height #6
}
\def\setpos(#1,#2){\xpos=#1 \ypos#2}

\def\settriparms[#1`#2`#3;#4]{\settripairparms[#1`#2`#3`1`1;#4]}%

\def\settripairparms[#1`#2`#3`#4`#5;#6]{%
\arrowtypea #1
\arrowtypeb #2
\arrowtypec #3
\arrowtyped #4
\arrowtypee #5
\width #6
\height #6
}

\def\resetparms{\settripairparms[1`1`1`1`1;500]\width 500}

\resetparms

\def\mvector(#1,#2)#3{
\put(0,0){\vector(#1,#2){#3}}%
\put(0,0){\vector(#1,#2){26}}%
}
\def\evector(#1,#2)#3{{
\arrowlength #3
\put(0,0){\vector(#1,#2){\arrowlength}}%
\advance \arrowlength by-30
\put(0,0){\vector(#1,#2){\arrowlength}}%
}}

\def\horsize#1#2{%
\settowidth{\tempdimen}{$#2$}%
#1=\tempdimen
\divide #1 by\unitlength
}

\def\vertsize#1#2{%
\settoheight{\tempdimen}{$#2$}%
#1=\tempdimen
\settodepth{\tempdimen}{$#2$}%
\advance #1 by\tempdimen
\divide #1 by\unitlength
}

\def\putvector(#1,#2)(#3,#4)#5#6{{%
\ifnum3<\arrowtype
\putdashvector(#1,#2)(#3,#4)#5\arrowtype
\else
\ifnum\arrowtype<-3
\putdashvector(#1,#2)(#3,#4)#5\arrowtype
\else
\xpos=#1
\ypos=#2
\run=#3
\rise=#4
\arrowlength=#5
\ifnum \arrowtype<0
    \ifnum \run=0
        \advance \ypos by-\arrowlength
    \else
        \tempcounta \arrowlength
        \multiply \tempcounta by\rise
        \divide \tempcounta by\run
        \ifnum\run>0
            \advance \xpos by\arrowlength
            \advance \ypos by\tempcounta
        \else
            \advance \xpos by-\arrowlength
            \advance \ypos by-\tempcounta
        \fi
    \fi
    \multiply \arrowtype by-1
    \multiply \rise by-1
    \multiply \run by-1
\fi
\ifcase \arrowtype
\or \put(\xpos,\ypos){\vector(\run,\rise){\arrowlength}}%
\or \put(\xpos,\ypos){\mvector(\run,\rise)\arrowlength}%
\or \put(\xpos,\ypos){\evector(\run,\rise){\arrowlength}}%
\fi\fi\fi
}}

\def\putsplitvector(#1,#2)#3#4{
\xpos #1
\ypos #2
\arrowtype #4
\halflength #3
\arrowlength #3
\gap 140
\advance \halflength by-\gap
\divide \halflength by2
\ifnum\arrowtype>0
   \ifcase \arrowtype
   \or \put(\xpos,\ypos){\line(0,-1){\halflength}}%
       \advance\ypos by-\halflength
       \advance\ypos by-\gap
       \put(\xpos,\ypos){\vector(0,-1){\halflength}}%
   \or \put(\xpos,\ypos){\line(0,-1)\halflength}%
       \put(\xpos,\ypos){\vector(0,-1)3}%
       \advance\ypos by-\halflength
       \advance\ypos by-\gap
       \put(\xpos,\ypos){\vector(0,-1){\halflength}}%
   \or \put(\xpos,\ypos){\line(0,-1)\halflength}%
       \advance\ypos by-\halflength
       \advance\ypos by-\gap
       \put(\xpos,\ypos){\evector(0,-1){\halflength}}%
   \fi
\else \arrowtype=-\arrowtype
   \ifcase\arrowtype
   \or \advance \ypos by-\arrowlength
       \put(\xpos,\ypos){\line(0,1){\halflength}}%
       \advance\ypos by\halflength
       \advance\ypos by\gap
       \put(\xpos,\ypos){\vector(0,1){\halflength}}%
   \or \advance \ypos by-\arrowlength
       \put(\xpos,\ypos){\line(0,1)\halflength}%
       \put(\xpos,\ypos){\vector(0,1)3}%
       \advance\ypos by\halflength
       \advance\ypos by\gap
       \put(\xpos,\ypos){\vector(0,1){\halflength}}%
   \or \advance \ypos by-\arrowlength
       \put(\xpos,\ypos){\line(0,1)\halflength}%
       \advance\ypos by\halflength
       \advance\ypos by\gap
       \put(\xpos,\ypos){\evector(0,1){\halflength}}%
   \fi
\fi
}

\def\putmorphism(#1)(#2,#3)[#4`#5`#6]#7#8#9{{%
\run #2
\rise #3
\ifnum\rise=0
  \puthmorphism(#1)[#4`#5`#6]{#7}{#8}#9%
\else\ifnum\run=0
  \putvmorphism(#1)[#4`#5`#6]{#7}{#8}#9%
\else
\setpos(#1)%
\arrowlength #7
\arrowtype #8
\ifnum\run=0
\else\ifnum\rise=0
\else
\ifnum\run>0
    \coefa=1
\else
   \coefa=-1
\fi
\ifnum\arrowtype>0
   \coefb=0
   \coefc=-1
\else
   \coefb=\coefa
   \coefc=1
   \arrowtype=-\arrowtype
\fi
\width=2
\multiply \width by\run
\divide \width by\rise
\ifnum \width<0  \width=-\width\fi
\advance\width by60
\if l#9 \width=-\width\fi
\putbox(\xpos,\ypos){#4}
{\multiply \coefa by\arrowlength
\advance\xpos by\coefa
\multiply \coefa by\rise
\divide \coefa by\run
\advance \ypos by\coefa
\putbox(\xpos,\ypos){#5} }%
{\multiply \coefa by\arrowlength
\divide \coefa by2
\advance \xpos by\coefa
\advance \xpos by\width
\multiply \coefa by\rise
\divide \coefa by\run
\advance \ypos by\coefa
\if l#9%
   \putrbox(\xpos,\ypos){#6}%
\else\if r#9%
   \putlbox(\xpos,\ypos){#6}%
\fi\fi }%
{\multiply \rise by-\coefc
\multiply \run by-\coefc
\multiply \coefb by\arrowlength
\advance \xpos by\coefb
\multiply \coefb by\rise
\divide \coefb by\run
\advance \ypos by\coefb
\multiply \coefc by70
\advance \ypos by\coefc
\multiply \coefc by\run
\divide \coefc by\rise
\advance \xpos by\coefc
\multiply \coefa by140
\multiply \coefa by\run
\divide \coefa by\rise
\advance \arrowlength by\coefa
\ifcase\arrowtype
\or \put(\xpos,\ypos){\vector(\run,\rise){\arrowlength}}%
\or \put(\xpos,\ypos){\mvector(\run,\rise){\arrowlength}}%
\or \put(\xpos,\ypos){\evector(\run,\rise){\arrowlength}}%
\fi}\fi\fi\fi\fi}}

\newcount\numbdashes \newcount\lengthdash \newcount\increment

\def\howmanydashes{
\numbdashes=\arrowlength \lengthdash=40
\divide\numbdashes by \lengthdash
\lengthdash=\arrowlength
\divide\lengthdash by \numbdashes
\increment=\lengthdash
\multiply\lengthdash by 3
\divide\lengthdash by 5
}

\def\putdashvector(#1)(#2,#3)#4#5{%
\ifnum#3=0 \putdashhvector(#1){#4}#5
\else
\ifnum#2=0
\putdashvvector(#1){#4}#5\fi\fi}

\def\putdashhvector(#1,#2)#3#4{{%
\arrowlength=#3 \howmanydashes
\multiput(#1,#2)(\increment,0){\numbdashes}%
{\vrule height .4pt width \lengthdash\unitlength}
\arrowtype=#4 \xpos=#1
\ifnum\arrowtype<0 \advance\arrowtype by 7 \fi
\ifcase\arrowtype
\or \advance\xpos by 10
    \put(\xpos,#2){\vector(-1,0){\lengthdash}}
    \advance\xpos by 40
    \put(\xpos,#2){\vector(-1,0){\lengthdash}}
\or \advance \xpos by 10
    \put(\xpos,#2){\vector(-1,0){\lengthdash}}
    \advance\xpos by  \arrowlength
    \advance\xpos by  -50
    \put(\xpos,#2){\vector(-1,0){\lengthdash}}
\or \advance\xpos by 10
    \put(\xpos,#2){\vector(-1,0){\lengthdash}}
\or \advance\xpos by \arrowlength
    \advance\xpos by -\lengthdash
    \put(\xpos,#2){\vector(1,0){\lengthdash}}
\or {\advance\xpos by 10
    \put(\xpos,#2){\vector(1,0){\lengthdash}}}
    \advance\xpos by \arrowlength
    \advance\xpos by -\lengthdash
    \put(\xpos,#2){\vector(1,0){\lengthdash}}
\or \advance\xpos by \arrowlength
    \advance\xpos by -\lengthdash
    \put(\xpos,#2){\vector(1,0){\lengthdash}}
    \advance\xpos by -40
    \put(\xpos,#2){\vector(1,0){\lengthdash}}
   \fi
}}

\def\putdashvvector(#1,#2)#3#4{{%
\arrowlength=#3 \howmanydashes
\ypos=#2 \advance\ypos by -\arrowlength
\multiput(#1,#2)(0,\increment){\numbdashes}%
    {\vrule width .4pt height \lengthdash\unitlength}
\arrowtype=#4 \ypos=#2
\ifnum\arrowtype<0 \advance\arrowtype by 7 \fi
\ifcase\arrowtype
\or \advance\ypos by \arrowlength \advance\ypos by -40
    \put(#1,\ypos){\vector(0,1){\lengthdash}}
    \advance\ypos by -40
    \put(#1,\ypos){\vector(0,1){\lengthdash}}
\or \advance\ypos by 10
    \put(#1,\ypos){\vector(0,1){\lengthdash}}
    \advance\ypos by \arrowlength \advance\ypos by -40
    \put(#1,\ypos){\vector(0,1){\lengthdash}}
\or \advance\ypos by \arrowlength \advance\ypos by -40
    \put(#1,\ypos){\vector(0,1){\lengthdash}}
\or \advance\ypos by 10
    \put(#1,\ypos){\vector(0,-1){\lengthdash}}
\or \advance\ypos by 10
    \put(#1,\ypos){\vector(0,-1){\lengthdash}}
    \advance\ypos by \arrowlength \advance\ypos by -40
    \put(#1,\ypos){\vector(0,-1){\lengthdash}}
\or \advance\ypos by 10
    \put(#1,\ypos){\vector(0,-1){\lengthdash}}
    \advance\ypos by 40
    \put(#1,\ypos){\vector(0,-1){\lengthdash}}
\fi
}}

\def\puthmorphism(#1,#2)[#3`#4`#5]#6#7#8{{%
\xpos #1
\ypos #2
\width #6
\arrowlength #6
\arrowtype=#7
\putbox(\xpos,\ypos){#3\vphantom{#4}}%
{\advance \xpos by\arrowlength
\putbox(\xpos,\ypos){\vphantom{#3}#4}}%
\horsize{\tempcounta}{#3}%
\horsize{\tempcountb}{#4}%
\divide \tempcounta by2
\divide \tempcountb by2
\advance \tempcounta by30
\advance \tempcountb by30
\advance \xpos by\tempcounta
\advance \arrowlength by-\tempcounta
\advance \arrowlength by-\tempcountb
\putvector(\xpos,\ypos)(1,0)\arrowlength\arrowtype
\divide \arrowlength by2
\advance \xpos by\arrowlength
\vertsize{\tempcounta}{#5}%
\divide\tempcounta by2
\advance \tempcounta by20
\if a#8 %
   \advance \ypos by\tempcounta
   \putbox(\xpos,\ypos){#5}%
\else
   \advance \ypos by-\tempcounta
   \putbox(\xpos,\ypos){#5}%
\fi}}

\def\putvmorphism(#1,#2)[#3`#4`#5]#6#7#8{{%
\xpos #1
\ypos #2
\arrowlength #6
\arrowtype #7
\settowidth{\xlen}{$#5$}%
\putbox(\xpos,\ypos){#3}%
{\advance \ypos by-\arrowlength
\putbox(\xpos,\ypos){#4}}%
{\advance\arrowlength by-140
\advance \ypos by-70
\ifdim\xlen>0pt
   \if m#8%
      \putsplitvector(\xpos,\ypos)\arrowlength\arrowtype
   \else
   \putvector(\xpos,\ypos)(0,-1)\arrowlength\arrowtype
   \fi
\else
   \putvector(\xpos,\ypos)(0,-1)\arrowlength\arrowtype
\fi}%
\ifdim\xlen>0pt
   \divide \arrowlength by2
   \advance\ypos by-\arrowlength
   \if l#8%
      \advance \xpos by-40
      \putrbox(\xpos,\ypos){#5}%
   \else\if r#8%
      \advance \xpos by40
      \putlbox(\xpos,\ypos){#5}%
   \else
      \putbox(\xpos,\ypos){#5}%
   \fi\fi
\fi
}}

\def\putsquarep<#1>(#2)[#3;#4`#5`#6`#7]{{%
\setsqparms[#1]%
\setpos(#2)%
\settokens[#3]%
\puthmorphism(\xpos,\ypos)[\tokenc`\tokend`{#7}]{\width}{\arrowtyped}b%
\advance\ypos by \height
\puthmorphism(\xpos,\ypos)[\tokena`\tokenb`{#4}]{\width}{\arrowtypea}a%
\putvmorphism(\xpos,\ypos)[``{#5}]{\height}{\arrowtypeb}l%
\advance\xpos by \width
\putvmorphism(\xpos,\ypos)[``{#6}]{\height}{\arrowtypec}r%
}}

\def\putsquare{\@ifnextchar <{\putsquarep}{\putsquarep%
   <\arrowtypea`\arrowtypeb`\arrowtypec`\arrowtyped;\width`\height>}}
\def\square{\@ifnextchar< {\squarep}{\squarep
   <\arrowtypea`\arrowtypeb`\arrowtypec`\arrowtyped;\width`\height>}}
\def\squarep<#1>[#2`#3`#4`#5;#6`#7`#8`#9]{{
\setsqparms[#1]
\diagram
\putsquarep<\arrowtypea`\arrowtypeb`\arrowtypec`
\arrowtyped;\width`\height>
(0,0)[#2`#3`#4`{#5};#6`#7`#8`{#9}]
\enddiagram
}}                                                 
\def\putptrianglep<#1>(#2,#3)[#4`#5`#6;#7`#8`#9]{{%
\settriparms[#1]%
\xpos=#2 \ypos=#3
\advance\ypos by \height
\puthmorphism(\xpos,\ypos)[#4`#5`{#7}]{\height}{\arrowtypea}a%
\putvmorphism(\xpos,\ypos)[`#6`{#8}]{\height}{\arrowtypeb}l%
\advance\xpos by\height
\putmorphism(\xpos,\ypos)(-1,-1)[``{#9}]{\height}{\arrowtypec}r%
}}

\def\putptriangle{\@ifnextchar <{\putptrianglep}{\putptrianglep
   <\arrowtypea`\arrowtypeb`\arrowtypec;\height>}}
\def\ptriangle{\@ifnextchar <{\ptrianglep}{\ptrianglep
   <\arrowtypea`\arrowtypeb`\arrowtypec;\height>}}
\def\ptrianglep<#1>[#2`#3`#4;#5`#6`#7]{{
\settriparms[#1]
\diagram
\putptrianglep<\arrowtypea`\arrowtypeb`
\arrowtypec;\height>
(0,0)[#2`#3`#4;#5`#6`{#7}]
\enddiagram
}}                                            

\def\putqtrianglep<#1>(#2,#3)[#4`#5`#6;#7`#8`#9]{{%
\settriparms[#1]%
\xpos=#2 \ypos=#3
\advance\ypos by\height
\puthmorphism(\xpos,\ypos)[#4`#5`{#7}]{\height}{\arrowtypea}a%
\putmorphism(\xpos,\ypos)(1,-1)[``{#8}]{\height}{\arrowtypeb}l%
\advance\xpos by\height
\putvmorphism(\xpos,\ypos)[`#6`{#9}]{\height}{\arrowtypec}r%
}}

\def\putqtriangle{\@ifnextchar <{\putqtrianglep}{\putqtrianglep
   <\arrowtypea`\arrowtypeb`\arrowtypec;\height>}}
\def\qtriangle{\@ifnextchar <{\qtrianglep}{\qtrianglep
   <\arrowtypea`\arrowtypeb`\arrowtypec;\height>}}
\def\qtrianglep<#1>[#2`#3`#4;#5`#6`#7]{{
\settriparms[#1]
\width=\height                                
\diagram
\putqtrianglep<\arrowtypea`\arrowtypeb`
\arrowtypec;\height>
(0,0)[#2`#3`#4;#5`#6`{#7}]
\enddiagram
}}

\def\putdtrianglep<#1>(#2,#3)[#4`#5`#6;#7`#8`#9]{{%
\settriparms[#1]%
\xpos=#2 \ypos=#3
\puthmorphism(\xpos,\ypos)[#5`#6`{#9}]{\height}{\arrowtypec}b%
\advance\xpos by \height \advance\ypos by\height
\putmorphism(\xpos,\ypos)(-1,-1)[``{#7}]{\height}{\arrowtypea}l%
\putvmorphism(\xpos,\ypos)[#4``{#8}]{\height}{\arrowtypeb}r%
}}

\def\putdtriangle{\@ifnextchar <{\putdtrianglep}{\putdtrianglep
   <\arrowtypea`\arrowtypeb`\arrowtypec;\height>}}
\def\dtriangle{\@ifnextchar <{\dtrianglep}{\dtrianglep
   <\arrowtypea`\arrowtypeb`\arrowtypec;\height>}}
\def\dtrianglep<#1>[#2`#3`#4;#5`#6`#7]{{
\settriparms[#1]
\width=\height                                
\diagram
\putdtrianglep<\arrowtypea`\arrowtypeb`
\arrowtypec;\height>
(0,0)[#2`#3`#4;#5`#6`{#7}]
\enddiagram
}}

\def\putbtrianglep<#1>(#2,#3)[#4`#5`#6;#7`#8`#9]{{%
\settriparms[#1]%
\xpos=#2 \ypos=#3
\puthmorphism(\xpos,\ypos)[#5`#6`{#9}]{\height}{\arrowtypec}b%
\advance\ypos by\height
\putmorphism(\xpos,\ypos)(1,-1)[``{#8}]{\height}{\arrowtypeb}r%
\putvmorphism(\xpos,\ypos)[#4``{#7}]{\height}{\arrowtypea}l%
}}

\def\putbtriangle{\@ifnextchar <{\putbtrianglep}{\putbtrianglep
   <\arrowtypea`\arrowtypeb`\arrowtypec;\height>}}
\def\btriangle{\@ifnextchar <{\btrianglep}{\btrianglep
   <\arrowtypea`\arrowtypeb`\arrowtypec;\height>}}
\def\btrianglep<#1>[#2`#3`#4;#5`#6`#7]{{
\settriparms[#1]
\width=\height                               
\diagram
\putbtrianglep<\arrowtypea`\arrowtypeb`
\arrowtypec;\height>
(0,0)[#2`#3`#4;#5`#6`{#7}]
\enddiagram
}}

\def\putAtrianglep<#1>(#2,#3)[#4`#5`#6;#7`#8`#9]{{%
\settriparms[#1]%
\xpos=#2 \ypos=#3
{\multiply \height by2
\puthmorphism(\xpos,\ypos)[#5`#6`{#9}]{\height}{\arrowtypec}b}%
\advance\xpos by\height \advance\ypos by\height
\putmorphism(\xpos,\ypos)(-1,-1)[#4``{#7}]{\height}{\arrowtypea}l%
\putmorphism(\xpos,\ypos)(1,-1)[``{#8}]{\height}{\arrowtypeb}r%
}}

\def\putAtriangle{\@ifnextchar <{\putAtrianglep}{\putAtrianglep
   <\arrowtypea`\arrowtypeb`\arrowtypec;\height>}}
\def\Atriangle{\@ifnextchar <{\Atrianglep}{\Atrianglep
   <\arrowtypea`\arrowtypeb`\arrowtypec;\height>}}
\def\Atrianglep<#1>[#2`#3`#4;#5`#6`#7]{{
\settriparms[#1]
\width=\height                                     
\diagram
\putAtrianglep<\arrowtypea`\arrowtypeb`
\arrowtypec;\height>
(0,0)[#2`#3`#4;#5`#6`{#7}]
\enddiagram
}}

\def\putAtrianglepairp<#1>(#2)[#3;#4`#5`#6`#7`#8]{{%
\settripairparms[#1]%
\setpos(#2)%
\settokens[#3]%
\puthmorphism(\xpos,\ypos)[\tokenb`\tokenc`{#7}]{\height}{\arrowtyped}b%
\advance\xpos by\height
\puthmorphism(\xpos,\ypos)[\phantom{\tokenc}`\tokend`{#8}]%
{\height}{\arrowtypee}b%
\advance\ypos by\height
\putmorphism(\xpos,\ypos)(-1,-1)[\tokena``{#4}]{\height}{\arrowtypea}l%
\putvmorphism(\xpos,\ypos)[``{#5}]{\height}{\arrowtypeb}m%
\putmorphism(\xpos,\ypos)(1,-1)[``{#6}]{\height}{\arrowtypec}r%
}}

\def\putAtrianglepair{\@ifnextchar <{\putAtrianglepairp}{\putAtrianglepairp%
   <\arrowtypea`\arrowtypeb`\arrowtypec`\arrowtyped`\arrowtypee;\height>}}
\def\Atrianglepair{\@ifnextchar <{\Atrianglepairp}{\Atrianglepairp%
   <\arrowtypea`\arrowtypeb`\arrowtypec`\arrowtyped`\arrowtypee;\height>}}

\def\Atrianglepairp<#1>[#2;#3`#4`#5`#6`#7]{{
\settripairparms[#1]
\settokens[#2]
\width=\height                                
\diagram
\putAtrianglepairp                            
<\arrowtypea`\arrowtypeb`\arrowtypec`
\arrowtyped`\arrowtypee;\height>
(0,0)[{#2};#3`#4`#5`#6`{#7}]
\enddiagram
}}

\def\putVtrianglep<#1>(#2,#3)[#4`#5`#6;#7`#8`#9]{{%
\settriparms[#1]%
\xpos=#2 \ypos=#3
\advance\ypos by\height
{\multiply\height by2
\puthmorphism(\xpos,\ypos)[#4`#5`{#7}]{\height}{\arrowtypea}a}%
\putmorphism(\xpos,\ypos)(1,-1)[`#6`{#8}]{\height}{\arrowtypeb}l%
\advance\xpos by\height
\advance\xpos by\height
\putmorphism(\xpos,\ypos)(-1,-1)[``{#9}]{\height}{\arrowtypec}r%
}}

\def\putVtriangle{\@ifnextchar <{\putVtrianglep}{\putVtrianglep
   <\arrowtypea`\arrowtypeb`\arrowtypec;\height>}}
\def\Vtriangle{\@ifnextchar <{\Vtrianglep}{\Vtrianglep
   <\arrowtypea`\arrowtypeb`\arrowtypec;\height>}}
\def\Vtrianglep<#1>[#2`#3`#4;#5`#6`#7]{{
\settriparms[#1]
\width=\height                                 
\diagram
\putVtrianglep<\arrowtypea`\arrowtypeb`
\arrowtypec;\height>
(0,0)[#2`#3`#4;#5`#6`{#7}]
\enddiagram
}}

\def\putVtrianglepairp<#1>(#2)[#3;#4`#5`#6`#7`#8]{{
\settripairparms[#1]%
\setpos(#2)%
\settokens[#3]%
\advance\ypos by\height
\putmorphism(\xpos,\ypos)(1,-1)[`\tokend`{#6}]{\height}{\arrowtypec}l%
\puthmorphism(\xpos,\ypos)[\tokena`\tokenb`{#4}]{\height}{\arrowtypea}a%
\advance\xpos by\height
\puthmorphism(\xpos,\ypos)[\phantom{\tokenb}`\tokenc`{#5}]%
{\height}{\arrowtypeb}a%
\putvmorphism(\xpos,\ypos)[``{#7}]{\height}{\arrowtyped}m%
\advance\xpos by\height
\putmorphism(\xpos,\ypos)(-1,-1)[``{#8}]{\height}{\arrowtypee}r%
}}

\def\putVtrianglepair{\@ifnextchar <{\putVtrianglepairp}{\putVtrianglepairp%
    <\arrowtypea`\arrowtypeb`\arrowtypec`\arrowtyped`\arrowtypee;\height>}}
\def\Vtrianglepair{\@ifnextchar <{\Vtrianglepairp}{\Vtrianglepairp%
    <\arrowtypea`\arrowtypeb`\arrowtypec`\arrowtyped`\arrowtypee;\height>}}
\def\Vtrianglepairp<#1>[#2;#3`#4`#5`#6`#7]{{
\settripairparms[#1]
\settokens[#2]
\diagram
\putVtrianglepairp                             
<\arrowtypea`\arrowtypeb`\arrowtypec`
\arrowtyped`\arrowtypee;\height>
(0,0)[{#2};#3`#4`#5`#6`{#7}]
\enddiagram
}}

\def\putCtrianglep<#1>(#2,#3)[#4`#5`#6;#7`#8`#9]{{%
\settriparms[#1]%
\xpos=#2 \ypos=#3
\advance\ypos by\height
\putmorphism(\xpos,\ypos)(1,-1)[``{#9}]{\height}{\arrowtypec}l%
\advance\xpos by\height
\advance\ypos by\height
\putmorphism(\xpos,\ypos)(-1,-1)[#4`#5`{#7}]{\height}{\arrowtypea}l%
{\multiply\height by 2
\putvmorphism(\xpos,\ypos)[`#6`{#8}]{\height}{\arrowtypeb}r}%
}}

\def\putCtriangle{\@ifnextchar <{\putCtrianglep}{\putCtrianglep
    <\arrowtypea`\arrowtypeb`\arrowtypec;\height>}}
\def\Ctriangle{\@ifnextchar <{\Ctrianglep}{\Ctrianglep
    <\arrowtypea`\arrowtypeb`\arrowtypec;\height>}}
\def\Ctrianglep<#1>[#2`#3`#4;#5`#6`#7]{{
\settriparms[#1]
\width=\height                               
\diagram
\putCtrianglep<\arrowtypea`\arrowtypeb`
\arrowtypec;\height>
(0,0)[#2`#3`#4;#5`#6`{#7}]
\enddiagram
}}                                           
\def\putDtrianglep<#1>(#2,#3)[#4`#5`#6;#7`#8`#9]{{%
\settriparms[#1]%
\xpos=#2 \ypos=#3
\advance\xpos by\height \advance\ypos by\height
\putmorphism(\xpos,\ypos)(-1,-1)[``{#9}]{\height}{\arrowtypec}r%
\advance\xpos by-\height \advance\ypos by\height
\putmorphism(\xpos,\ypos)(1,-1)[`#5`{#8}]{\height}{\arrowtypeb}r%
{\multiply\height by 2
\putvmorphism(\xpos,\ypos)[#4`#6`{#7}]{\height}{\arrowtypea}l}%
}}

\def\putDtriangle{\@ifnextchar <{\putDtrianglep}{\putDtrianglep
    <\arrowtypea`\arrowtypeb`\arrowtypec;\height>}}
\def\Dtriangle{\@ifnextchar <{\Dtrianglep}{\Dtrianglep
   <\arrowtypea`\arrowtypeb`\arrowtypec;\height>}}
\def\Dtrianglep<#1>[#2`#3`#4;#5`#6`#7]{{
\settriparms[#1]
\width=\height                              
\diagram
\putDtrianglep<\arrowtypea`\arrowtypeb`
\arrowtypec;\height>
(0,0)[#2`#3`#4;#5`#6`{#7}]
\enddiagram
}}                                          
\def\setrecparms[#1`#2]{\width=#1 \height=#2}%

\def\recursep<#1`#2>[#3;#4`#5`#6`#7`#8]{{%
\width=#1 \height=#2
\settokens[#3]
\settowidth{\tempdimen}{$\tokena$}
\ifdim\tempdimen=0pt
  \savebox{\tempboxa}{\hbox{$\tokenb$}}%
  \savebox{\tempboxb}{\hbox{$\tokend$}}%
  \savebox{\tempboxc}{\hbox{$#6$}}%
\else
  \savebox{\tempboxa}{\hbox{$\hbox{$\tokena$}\times\hbox{$\tokenb$}$}}%
  \savebox{\tempboxb}{\hbox{$\hbox{$\tokena$}\times\hbox{$\tokend$}$}}%
  \savebox{\tempboxc}{\hbox{$\hbox{$\tokena$}\times\hbox{$#6$}$}}%
\fi
\ypos=\height
\divide\ypos by 2
\xpos=\ypos
\advance\xpos by \width
\bfig
\putCtrianglep<-1`1`1;\ypos>(0,0)[`\tokenc`;#5`#6`{#7}]%
\puthmorphism(\ypos,0)[\tokend`\usebox{\tempboxb}`{#8}]{\width}{-1}b%
\puthmorphism(\ypos,\height)[\tokenb`\usebox{\tempboxa}`{#4}]{\width}{-1}a%
\advance\ypos by \width
\putvmorphism(\ypos,\height)[``\usebox{\tempboxc}]{\height}1r%
\efig
}}

\def\recurse{\@ifnextchar <{\recursep}{\recursep<\width`\height>}}

\def\puttwohmorphisms(#1,#2)[#3`#4;#5`#6]#7#8#9{{%
%
\puthmorphism(#1,#2)[#3`#4`]{#7}0a
\ypos=#2
\advance\ypos by 20
\puthmorphism(#1,\ypos)[\phantom{#3}`\phantom{#4}`#5]{#7}{#8}a
\advance\ypos by -40
\puthmorphism(#1,\ypos)[\phantom{#3}`\phantom{#4}`#6]{#7}{#9}b
}}

\def\puttwovmorphisms(#1,#2)[#3`#4;#5`#6]#7#8#9{{%
%
%
\putvmorphism(#1,#2)[#3`#4`]{#7}0a
\xpos=#1
\advance\xpos by -20
\putvmorphism(\xpos,#2)[\phantom{#3}`\phantom{#4}`#5]{#7}{#8}l
\advance\xpos by 40
\putvmorphism(\xpos,#2)[\phantom{#3}`\phantom{#4}`#6]{#7}{#9}r
}}

\def\puthcoequalizer(#1)[#2`#3`#4;#5`#6`#7]#8#9{{%
%
\setpos(#1)%
\puttwohmorphisms(\xpos,\ypos)[#2`#3;#5`#6]{#8}11%
\advance\xpos by #8
\puthmorphism(\xpos,\ypos)[\phantom{#3}`#4`#7]{#8}1{#9}
}}

\def\putvcoequalizer(#1)[#2`#3`#4;#5`#6`#7]#8#9{{%
%
%
\setpos(#1)%
\puttwovmorphisms(\xpos,\ypos)[#2`#3;#5`#6]{#8}11%
\advance\ypos by -#8
\putvmorphism(\xpos,\ypos)[\phantom{#3}`#4`#7]{#8}1{#9}
}}

\def\putthreehmorphisms(#1)[#2`#3;#4`#5`#6]#7(#8)#9{{%
\setpos(#1) \settypes(#8)
\if a#9 %
     \vertsize{\tempcounta}{#5}%
     \vertsize{\tempcountb}{#6}%
     \ifnum \tempcounta<\tempcountb \tempcounta=\tempcountb \fi
\else
     \vertsize{\tempcounta}{#4}%
     \vertsize{\tempcountb}{#5}%
     \ifnum \tempcounta<\tempcountb \tempcounta=\tempcountb \fi
\fi
\advance \tempcounta by 60
\puthmorphism(\xpos,\ypos)[#2`#3`#5]{#7}{\arrowtypeb}{#9}
\advance\ypos by \tempcounta
\puthmorphism(\xpos,\ypos)[\phantom{#2}`\phantom{#3}`#4]{#7}{\arrowtypea}{#9}
\advance\ypos by -\tempcounta \advance\ypos by -\tempcounta
\puthmorphism(\xpos,\ypos)[\phantom{#2}`\phantom{#3}`#6]{#7}{\arrowtypec}{#9}
}}

\def\setarrowtoks[#1`#2`#3`#4`#5`#6]{%
\def\toka{#1}
\def\tokb{#2}
\def\tokc{#3}
\def\tokd{#4}
\def\toke{#5}
\def\tokf{#6}
}
\def\hex{\@ifnextchar <{\hexp}{\hexp<1000`400>}}
\def\hexp<#1`#2>[#3`#4`#5`#6`#7`#8;#9]{%
\setarrowtoks[#9]
\yext=#2 \advance \yext by #2
\xext=#1 \advance\xext by \yext
\bfig
\putCtriangle<-1`0`1;#2>(0,0)[`#5`;\tokb``\tokd]
\xext=#1 \yext=#2 \advance \yext by #2
\putsquare<1`0`0`1;\xext`\yext>(#2,0)[#3`#4`#7`#8;\toka```\tokf]
\advance \xext by #2
\putDtriangle<0`1`-1;#2>(\xext,0)[`#6`;`\tokc`\toke]
\efig
}


\newtheorem{observation}{Remark}[section]
\newtheorem{lem}[observation]{Lemma}  
\newtheorem{thm}[observation]{Theorem}
\newtheorem{dfn}[observation]{Definition}
\newtheorem{exm}[observation]{Example}
\newtheorem{rem}[observation]{Remark}
\newtheorem{nte}[observation]{Notation}
\newtheorem{pro}[observation]{Proposition} 
\newtheorem{cor}[observation]{Corollary} 
\newtheorem{nada}[observation]{}  
\newcommand{\ph}[2]{\begin{#1} \label{#2}} 

\newcommand{\proof}{\noindent {\bf Proof} \hspace{.1in}}
\newcommand{\proofsketch}{\noindent {\bf Proof Sketch} \hspace{.1in}}
\newcommand\monadext[1]{ {#1} ^ \# }   
\newcommand\qed{$\hfill{\Box}$  } 
\newcommand\alg[2]{\ensuremath{\mathbf{#1}^{\mbox{\small \textbf{#2}}}}} 
\newcommand\kl[2]{\ensuremath{\mathbf{#1}_{\mbox{\small \textbf{#2}}}}} 
\newcommand\nats{\hbox{$I \kern - .38em N$}} 
\newcommand\ints{\hbox{$Z \kern - .65em Z$}} 
\newcommand\ssub[1]{_{{}_{#1}}}  
\newcommand\pointyPT[1]{\ensuremath{< \hspace{-.05in} #1 \hspace{-.05in} > \hspace{-.02in}}}
\newcommand\Lrightarrow{\hbox to 20pt{\rightarrowfill}} 
\newcommand\LOrightarrow{\hbox to 30pt{\rightarrowfill}}
\newcommand\LONrightarrow{\hbox to 45pt{\rightarrowfill}}
\newcommand\LONGrightarrow{\hbox to 60pt{\rightarrowfill}}
\newcommand\LONGErightarrow{\hbox to 75pt{\rightarrowfill}}
\newcommand\LONGERrightarrow{\hbox to 90pt{\rightarrowfill}}
\newcommand\ssr[1]{\stackrel{#1}{\Lrightarrow}} 
\newcommand\csr[1]{\stackrel{#1}{\LOrightarrow}} 
\newcommand\bsr[1]{\stackrel{#1}{\LONrightarrow}} 
\newcommand\Bsr[1]{\stackrel{#1}{\LONGrightarrow}} 

\newcommand\sets{\ensuremath{ \EuScript{S}}}   
\newcommand\cm{\mathbf{K} \circ_\sigma \mathbf{H}}  

\newcommand\bimapsto{\leftarrow \kern -.15em \mapsto}  
\newcommand\double[1]{#1 \mbox{\hspace{-0.2em}} #1 }  
\newcommand\cons{\hbox{+ \kern -.9em +}} 

\label{firstpage}
\maketitle

\begin{abstract}
Over the past two decades the notion of a strong monad has found wide applicability in computing. Arising out of a need to interpret products in computational and semantic settings, different approaches to this concept have arisen. In this paper we introduce and investigate the connections between these approaches and also relate the results to monad composition. We also introduce new methods for checking and using the required laws associated with such compositions, as well as provide examples illustrating problems and issues that arise.   
\end{abstract}

\section{Introduction}
The notion of a strong monad arises in numerous applications including programming language design and semantics, monadic interpreters, and building models of computation. The notion, originating with Kock\cite{articleB}, was originally defined for symmetric monoidal closed categories. Mulry and others exploited this idea to introduce specific kinds of strong monads to generate a wide variety of categorical computational models as well as an analysis of fixed point semantics\cite{articleM}.  In the work of Moggie and Wadler the notion of a strong monad was used to define the semantics of programming languages \cite{articleI}, \cite{articleL}. 

Different approaches to the notion of a strong monad have also arisen. From early on it was recognized that a strong monad, and particularly the special case of a commutative monad, was critical in providing a means of interpreting products in a Kleisli category, and a means of providing a functorial lifting of biproducts. More recently Manes and Mulry have focused attention on the idea that monads, rather than being called strong which might suggest that this is a uniquely defined property of a given monad, could be viewed as being equipped with a possibly non-unique strength of a given order. In  \cite{articleD}, the authors introduced the concept of a Kleisli strength of order $n$ for arbitrary  monads in an arbitrary symmetric monoidal category, demonstrated that the notion of Kleisli strength differs from Kock strength and provided examples demonstrating that Kleisli strength takes the form of added (non-unique) structure on a given monad, leading to the terminology of a monad equipped with a strength $(H,\Gamma)$. The authors further observed that the existence of Kleisli strength can be used to derive methods for generating (non-unique) composition of monads for a variety of monads including standard monadic data types such as lists, trees, exceptions, reader, writer and state monads.

Different ways of defining strength lead naturally to the question of describing more fully the connections between these definitions. For instance, the definition of strength found in some examples of strong monads used for monadic interpreters may not support products in the Kleisli category because they fail to satisfy the laws required of a Kleisli strength. A further complication arises from the observation that the original use of the notion by Kock assumed the monad itself was a monoidal functor, something most monads used in computing simply are not.

This paper provides details of the relationship between these varying notions of strength and suggests new nomenclature to distinguish between them. The paper assumes the reader is familiar with the general notions of category theory, and also depends heavily on \cite{articleD}. In section 2 we provide preliminaries such as defining the notion of a Kleisli strength, and include some examples and methods for generating it. As the examples illustrate, this notion is of independent interest and generalizes the prior notions of a commutative monad. 

Section 3 provides a brief introduction to classical commutative monads and Kock strength contrasting these concepts with that of general Kleisli strength. Also introduced is the notion of a prestrength construction arising from the existence of $map$, its associated category, as well as a wide variety of examples to help sort out the differences between the concepts.

In section 4 we restate and prove some of the previous results in Haskell syntax. In addition to providing a means of computing and using Kleisli strength and monad compositions in Haskell, it is hoped this approach may prove more intuitive and accessible to the reader used to working in the environment of functional programming.

\section{Preliminaries}
Let \textbf{C}, \textbf{D} be categories with functor $F: {\bf C} \rightarrow {\bf D}$,
$\mathbf{H} = (H, \mu, \eta)$ a monad in \textbf{C}, $\mathbf{K} = (K, \nu, \rho)$ a monad in \textbf{D}.  Note that we use categorical notation here.  We write $\kl{C}{H}$ for the Kleisli category of \textbf{H} with canonical functor  $\iota_\mathbf{H} : \mathbf{C} \rightarrow \kl{C}{H}$, $\iota_\mathbf{H} (X \ssr{f} Y) = X \ssr{f} Y \ssr{\eta_Y} HY$.  

A \textbf{Kleisli lifting} of functor $F$ is a functor $\overline{F}:~\kl{C}{H} \rightarrow \kl{D}{K}$ that lifts $F$ in the sense that the following diagram commutes.
\begin{center}
\setsqparms[1`-1`-1`1;500`500]
\square[\kl{C}{H}`\kl{D}{K}`\mathbf{C}`\mathbf{D};\overline{F}`\iota_\mathbf{H}`\iota_\mathbf{K}`F]
\end{center} 

Kleisli liftings are classified exactly by natural transformations $\lambda : FH \rightarrow KF$ satisfying 

\begin{center}
\xext=2000 \yext=700
\adjust[`\mu;`n;`Tn;`\mu]
\begin{picture}(\xext,\yext)(\xoff,\yoff)
\putmorphism(0,600)(1,0)[F`FH`F \eta]{975}1a
\putmorphism(1150,600)(1,0)[`FHH`F \mu]{850}{-1}a
\putmorphism(1050,-70)(1,0)[`KKF`\nu F ]{975}{-1}b
\putmorphism(975,600)(0,-1)[`KF`\lambda]{660}1r
\putmorphism(2015,600)(0,-1)[``\lambda H ]{330}1r
\putmorphism(2015,275)(0,-1)[KFH``K \lambda]{330}1r
\putmorphism(0,600)(4,-3)[``]{980}1a
\put(450,150){\makebox(0,0){$\rho F$}}
\put(600,350){\makebox(0,0){$(\overline{F} C)$}}
\put(1500,350){\makebox(0,0){$(\overline{F} D)$}}
\end{picture}
\end{center}
\vspace{.1in}

\medskip The transformation $\lambda$ is referred to as a \textbf{lifting transformation}. For instance, it is easy to see that given such a $\lambda$ and defining $\overline{F}$ by $\overline{F}(f:A \rightarrow HB) = \lambda \circ (F~f)$ results in a well-defined functor. It should be further pointed out that liftings (and thus lifting transformations) may not be unique for a fixed functor $F$, something we will demonstrate later. A simple example is the well known notion of a $\it monad~map: \kl{C}{H} \rightarrow \kl{C}{K}$ which corresponds to a Kleisli lift of the identity functor on \textbf{C}. This is a rather simple explanation of a far more subtle process. An elegant and more conceptual account of the precise correspondence and interaction between liftings and morphisms of monads can be found in \cite{articleJ}. 

While most of our discussion will take place in a cartesian closed category such as $\mathbf{Set}$, some of the following constructions and results are applicable to more general settings such as symmetric monoidal categories $\mathcal{V}$ with tensor $\otimes$ and tensor unit $I$. We will make occasional use of these generalizations here. 

We shall use the abbreviation $\times_n \: x$ for $x \times \cdots \times  x$ ($n$ times), where $x$ can be either an object or a map.  When $x$ is an object and $n=0$, $\times_n \: x \,=\,I$.  In general, $\times_1\:x = x$.  We also use the same symbol for the $n$-fold product functor, so that $\times_n \: (V_1,\ldots,V_n) = V_1 \times \cdots \times V_n$.

\section{Notions of Strength}

Let $\mathbf{K} = (K,\nu,\rho)$ be a monad on $\mathcal{V}$. For $n\ge 0$, consider the cartesian power category $\mathcal{V}^n$ with objects $(V_1,\ldots,V_n)$ and morphisms $(f_1,\ldots,f_n) : (V_1,\ldots,V_n) \rightarrow (W_1,\ldots,W_n)$.  \textbf{K} induces a monad $\mathbf{K^{(n)}}$ on $\mathcal{V}^n$ via $K^{(n)}(V_1,\ldots,V_n) = (KV_1,\ldots,KV_n)$, $\rho \ssub{(V_1,\ldots,V_n)} = (\rho \ssub{V_1},\ldots,\rho \ssub{V_n})$, $\nu \ssub{(V_1,\ldots,V_n)} = (\nu \ssub{V_1},\ldots,\nu \ssub{V_n})$.  We denote the resulting Kleisli category for $\mathbf{K^{(n)}}$ simply as $\mathcal{V}_{\mathbf{K}}^n$, also reflecting the fact that the Kleisli construction commutes with products in $\mathbf{Cat}$.  

\subsection{Pre-strength}

Many of the notions and examples found in the first two sections can also be found in \cite{articleD}, including the notion of $prestrength$.

\ph{dfn}{prnv}
Consider the category whose objects are pairs $(F,\Gamma^F)$ with $F : \mathcal{V} \rightarrow \mathcal{V}$ a functor and $\Gamma^F_{V_1 \cdots V_n} : FV_1 \times \cdots \times FV_n \rightarrow F(V_1 \times \cdots \times V_n)$ a natural transformation.  Such a $\Gamma^F$ is called a \textbf{pre-strength} on 
$F$ (we often drop the subscripts except when needed for emphasis).  A morphism $\alpha : (F,\Gamma^F) \rightarrow (G,\Gamma^G)$ is a natural transformation $\alpha : F \rightarrow G$ such that the following square commutes.
\begin{center}
\square<1`1`1`1;1700`600>[FV_1 \times \cdots \times FV_n`GV_1 \times \cdots \times GV_n`F(V_1 \times \cdots \times V_n)`G(V_1 \times \cdots \times V_n);\alpha \ssub{V_1} \times \cdots \times \alpha \ssub{V_n}`\Gamma^F \ssub{V_1\cdots V_n}`\Gamma^G \ssub{V_1\cdots V_n}`\alpha \ssub{V_1 \times \cdots \times V_n}] 
\end{center}  
\end{dfn}
When $\alpha$ is such a morphism we say $\alpha$ \textbf{preserves pre-strengths}.  It is obvious that this is a category under vertical composition of natural transformations, but it is also a 2-category under the horizontal composition of the endofunctor category of $\mathcal{V}$.

\ph{exm}{comprehensionssformpre-strength}
If $\mathbf{F} = \mathbf{L}^+$ is the non-empty list monad and one defines defines $\Gamma^F$ via comprehensions, namely $\Gamma^F(list1,list2) = [(a,b)|a \leftarrow list1, b \leftarrow list2]$, it is easy to check that 
$\Gamma^F$ is a natural transformation and forms a pre-strength of order 2.   We return to this example shortly.
\end{exm}

\subsection{Kleisli strength}

In \cite{articleE} the idea was promoted that in the context of programming language semantics, a Kleisli lifting provides a means of lifting a given functorial process $F:\mathbf{C} \rightarrow \mathbf{D}$ to the semantics residing in the corresponding Kleisli categories $\overline{F}:~\kl{C}{H} \rightarrow \kl{D}{K}$. The associated lifting transformation mediates this process. A special case of this corresponds to lifting the product bifunctor to the corresponding Kleisli category.  In short, if products are to be well defined in the Kleisli category $~\kl{C}{H}$ and capable of interpreting composition of functions for computational or programming semantics as described earlier, a Kleisli lifting of the bifunctor must exist. This has been generalized to the notion of Kleisli strength.

\ph{dfn}{Kleislistrength}
A \textbf{Kleisli strength} on $\mathbf{K}$ of order $n \geq 0$ is a natural transformation $\Gamma^n \ssub{V_1,\cdots,V_n} : KV_1 \times \cdots \times KV_n \rightarrow K(V_1 \times \cdots \times V_n)$ which classifies a Kleisli lift $\overline{\times_n} : \mathcal{V}_{\mathbf{K}}^n \rightarrow \mathcal{V}_{\mathbf{K}}$ of the $n$-fold product functor $\times_n : \mathcal{V}^n \rightarrow \mathcal{V}$.  Thus $\Gamma^n$ is a natural transformation satisfying $(\Gamma^n\,A)$ and $(\Gamma^n\,B)$ as shown in the diagram below. (Note $V$-~subscripts are dropped for readability sake.) 
\end{dfn}  

\begin{center}
\xext=3200 \yext=700
\adjust[`\mu;`n;`Tn;`\mu]
\begin{picture}(\xext,\yext)(\xoff,\yoff)
\putmorphism(0,600)(1,0)[V_1 \times \cdots \times V_n`KV_1 \times \cdots \times KV_n`\rho \times \cdots \times \rho]{1400}1a
\putmorphism(1850,600)(1,0)[`KKV_1 \times \cdots \times KKV_n`\nu \times \cdots \times \nu]{1200}{-1}a
\putmorphism(1800,-70)(1,0)[`KK(V_1 \times \cdots \times V_n)`\nu  ]{1200}{-1}b
\putmorphism(1400,600)(0,-1)[`K(V_1 \times \cdots \times V_n)`\Gamma^n]{660}1r
\putmorphism(3015,600)(0,-1)[``\Gamma^n_K ]{330}1r
\putmorphism(3015,275)(0,-1)[K(KV_1 \times \cdots \times KV_n)``K \Gamma^n]{330}1r
\putmorphism(0,600)(4,-3)[``]{980}1a
\put(450,125){\makebox(0,0){$\rho $}}
\put(800,300){\makebox(0,0){$(\Gamma^n \, A)$}}
\put(2200,300){\makebox(0,0){$(\Gamma^n \,B)$}}
\end{picture}
\end{center}
\vspace{.2in}

For $n=0$, $(\Gamma^0\,A)$ always exists and coincides with $\rho_I$. A Kleisli strength of order $1$ is the same thing as a monad map $\mathbf{K} \rightarrow \mathbf{K}$. We note that there is always at least one Kleisli strength of order 1, namely $\Gamma^1_V = id_{KV}$. Based on the discussion of section 2, a Kleisli strength of order $n$ is precisely a lifting transformation for $\times_n$.

\vspace{.2in}
We briefly describe a few examples of Kleisli strength. As indicated earlier, the category of sets and total functions will be denoted $\mathbf{Set}$.

\ph{exm}{revisKleislistrength}
Let $\mathbf{K} = \mathbf{L}$ be the list monad then the reverse transformation $rev : L \rightarrow L$ is a Kleisli strength of order 1 on \textbf{L}. Since identity is also of order one by the above remark, this  demonstrates that Kleisli strengths are not unique. 
\end{exm}

\ph{exm}{powersetKleislistrength}
Let $\mathbf{K} = \mathbf{P_0}$ be the finite power set monad.  Then $\Gamma^n(A_1,\ldots,A_n) = A_1 \times \cdots \times A_n$ is a Kleisli strength of order $n\ge 0$, where $A_i\in P_0V_i$ for $1\le i\le n$.
\end{exm}

\ph{exm}{exceptionsarestrongmonads}
Let $\mathbf{K}$ be the exceptions monad $MX = X + Exc$, with $Exc$ any nonempty set. If $a \in Exc$, for each $n\ge 0$ the monad admits a Kleisli strength of order n $\Gamma^n$, where    $\Gamma^n(x_1,...,x_n) =(x_1, \dots ,x_n)$, if all of the $x_i$ are in $X$, and equals $a$ otherwise. If $a \not= b$ where $b \in Exc$, then the new $\Gamma^n$ generated by $b$ differs from the one generated by $a$. We return to this example shortly.
\end{exm}

\ph{exm}{bottommonadstrength} Much as in the previous example, let $\mathbf{K}$ be the \lq$add~a~bottom$\rq $~$or lifting monad, $KA = A_\bot$ defined on $\mathbf{Dom}$, the category of (possibly bottomless) Scott domains and continuous maps \cite{articleG}. $\Gamma^n$ is a Kleisli strength of order n ($n \geq 1$) where $\Gamma^n (x_1 \dots x_n) = (x_1 \dots x_n)$  if all of the $x_i$ are in $X$, and equals $\bot$ otherwise. 
\end{exm}

\ph{exm}{exponentialisstrong} For a fixed $A$ consider the exponential or reader monad $\mathbf{M}$ where $MB = A \rightarrow B$.  $M$ has a Kleisli strength of order n where $\Gamma^n (f_1 \dots f_n) = \lambda a.(f_1(a) \dots f_n(a))$. 
\end{exm}

\ph{exm}{certificatemonadnotstrong} Consider the M-set or writer monad $K X = M \times X$, where $M$ is a commutative monoid, then $\Gamma^2((m_1,x),(m_2,y)) = (m_1m_2,x,y))$ forms a Kleisli strength of order 2.
\end{exm}

In the last example if $M$ is not commutative, then $\Gamma^2$ need not form a Kleisli strength of order 2. Likewise in the next example we see list comprehensions do not generally form Kleisli strengths.

\ph{exm}{comprehensionssnotstrong}
Let $\mathbf{K} = \mathbf{L}^+$ then the prestrength $\Gamma^K$ of Example \ref{comprehensionssformpre-strength} does not form a Kleisli strength of order 2. Working clockwise on the pair $([[1,2],[3]], [[5,6],[7]])$ in diagram $(\Gamma^n\,B)$, for instance, generates $[(1,4),(1,5),(1,6),...]$, while counterclockwise generates 
$[(1,4),(1,5),(2,4),...]$. 
\end{exm}

\ph{exm}{binarytreemonad} The \textbf{binary tree monad} $VX$ consists of binary trees whose values (from $X$) are located in its leaves. We denote an empty tree by $E$, a trivial tree (i.e. a leaf) with value $x$ by $L(x)$ and a tree consisting of left and right subtrees $v1$ and $v2$ by $N(v1,v2)$. If $\mathbf{V}^+ $ denotes the submonad of non-empty binary trees, then the map $\Gamma:V^+A \times V^+B \rightarrow V^+(A \times B)$ defined by $\Gamma(t1,t2) = L(a,b)$ where $L(a)$ and $L(b)$ are the leftmost nodes of trees $t1$ and $t2$ respectively forms a Kleisli strength of order 2 on $V^+$.
\end{exm}

\ph{pro}{Kleislistrengthnfrom2}
Let $\Gamma^2$ be a Kleisli strength on $\mathbf{K}$ of order 2, then $\Gamma^3 = \Gamma^2\circ(\Gamma^2 \times 1)$ is a Kleisli strength on $\mathbf{K}$ of order 3.  The same process produces $\Gamma^4$ from $\Gamma^3$, $\Gamma^5$ from $\Gamma^4$, $\ldots$ producing a Kleisli strength $\Gamma^n$ of order $n$ for all $n \geq 2$. In a similar fashion one can also define a Kleisli strength of order 3 via $\Gamma^3 = \Gamma^2 \circ (1 \times \Gamma^2)$ and similarly $\Gamma^4 = \Gamma^2 \circ (1 \times \Gamma^3)$  $\ldots$. These constructions generally disagree unless $\Gamma^2$ is \textbf{associative}, namely $\Gamma \circ (\Gamma \times 1) = \Gamma \circ (1 \times \Gamma)$. 
\end{pro}    

\ph{exm}{non-emptylistsarestrongmonads} Let $\mathbf{L}^+$ be the non-empty list monad. In contrast to Example \ref{comprehensionssnotstrong}, by using a different approach than comprehensions we can produce examples of Kleisli strength. Let $fst$ and $lst: \mathbf{L}^+ \rightarrow id$ choose the first (or respectively last) element of a non-empty list. Then $\eta \circ (fst \times fst)$ and $\eta \circ (lst \times lst)$ are associative Kleisli strengths of order 2.
\end{exm}

\subsection{Prestrength Construction}

As the prior sections illustrate, the notion of Kleisli strength consists of additional (non-unique) structure $(\Gamma^n)$ imposed on the monad. In particular, no assumption was made of the existence of an internal map structure $st:(A \rightarrow B) \rightarrow (MA \rightarrow MB)$ for a monad $M$. The underlying category need not even be closed. This is a common assumption in functional programming, indeed it is already built into the specification of the functor class in Haskell, where it is denoted $fmap$. The idea of a monad in a category equipped with a map $st:B^A \rightarrow (MB)^{MA}$ had already been defined much earlier by Kock \cite{articleB} who called it a {\it strong monad}. Building the composition $\lambda a. st(\lambda b.(a,b)): A \rightarrow (MB \rightarrow M(A \times B))$, and taking adjoints generates the map we denote $rst:A \times MB \rightarrow M(A \times B))$, which is often referred to in the functional programming language literature as "the monad strength" for $M$ (and often denoted simply by $t$). In a similar fashion one can generate the map $lst:MA \times B \rightarrow M(A \times B))$. This leads in turn to the compositions

\medskip
$lst\Gamma_{AB}= rst^{\#} \circ lst: MA \times MB \rightarrow M(A \times B)$\\

$rst\Gamma_{AB}= lst^{\#} \circ rst: MA \times MB \rightarrow M(A \times B)$\\

\noindent where $f^{\#} a = (\mu \circ Mf)(a)$, using the standard ${\#}$ notation first introduced by Manes \cite{bookA}. These two compositions can easily be checked for naturality and form what we denote as the {\it left and right pre-strength constructions}, $lst\Gamma_{AB}, ~rst\Gamma_{AB}$, for the given monad.  These derived pre-strengths are generally not equal nor are they always examples of Kleisli strength as we illustrate next. In light of this last observation, it makes sense to consider to what extent prestrength constructions support products in a semantic setting.

\medskip
For the remainder of the paper whenever we have a monad $M$ on a cartesian closed category $\textbf{C}$ equipped with such a $st$ we will denote $(\textbf{C},M,st)$ as a {\bf $st$-monad}.

\ph{exm}{lGammanotaKleislistrengthI} If $\mathbf{L}^+$ is the non-empty list monad, the derived left prestrength $lst\Gamma$ is exactly the comprehension $\Gamma^F$ of Example \ref{comprehensionssformpre-strength}. Example \ref{comprehensionssnotstrong} provides an easy example where $(\Gamma^n\,B)$ fails for $n = 2$. Similarly $rst\Gamma$, which is the reverse comprehension, does not agree with $lst\Gamma$ nor does it satisfy $(\Gamma^n\,B)$. Thus neither is a Kleisli strength.
\end{exm}

In \cite{articleB}, Anders Kock assumed $\mathcal{V}$ to be a symmetric monoidal \textit{closed} category,  required $K$ to be a monoidal monad and $\nu$ and $\rho$ to be $\mathcal{V}$-natural. Under such assumptions Kock defined $K$ to be commutative if  $lst\Gamma = rst\Gamma$. Thus the previous example illustrates that even with a weaker notion of commutative monad in a cartesian closed category, $K$ need not be commutative. For the remainder of the paper we will let the term {\it commutative monad} refer to this weaker notion, namely a $st$-monad \ $(\textbf{C},M,st)$ in a cartesian closed category where the derived left and right prestrengths agree, $lst\Gamma = rst\Gamma$. This agrees with the usual use of this term, found for instance in \cite{articleI}.

\subsection{Monad semantics, cartesian products, and Kleisli Strength}

In \cite{articleL}, Wadler introduced the notion of a monadic interpreter. This application, as well as others in computation and programming semantics, depends on interpreting terms in the Kleisli category for a monad. In the case of monadic interpreters there is a core interpretation which applies to all monads with small modifications made for any particular monad. Notably, the example interpreters found there and elsewhere lack a product construct despite its obvious role in both programming and computation. This is no doubt due to the fact that such interpreters would have to satisfy certain laws ensuring that composition is well defined with regard to that construct in the associated Kleisli categories. As emphasized earler, such laws are satisfied precisely when the monad is equipped with a Kleisli strength.

Suppose we are given terms $t_1,t_2$ where each $t_i$ is interpreted $t_i:A \rightarrow MB_i$ for some monad $M$. The resulting term $(t_1,t_2)$ should have type $A \rightarrow M(B_1 \times B_2)$. This requires the existence of a prestrength $\Gamma:MA \times MB \rightarrow M(A \times B)$  with the resulting composition $\Gamma \circ <t_1,t_2>$. $\Gamma$ must be a Kleisli strength, a general prestrength will not do, nor even a derived pre-strength, in order for function composition to behave properly. It does not require that $\Gamma$ be commutative however. We illustrate this point with the next example. 

Let  $f:A \rightarrow B$, $g:A \rightarrow C$, $f~':B \rightarrow B~'$, $g~':C \rightarrow C~'$ be four arbitrary functions respectively. In any cartesian category \textbf{C}, it follows from the definition of product that the following two compositions (of type $A \rightarrow (B~' \times C~')$) must agree:

1) $comp1 = <f~'\circ f, g~'\circ g>:A \rightarrow (B~' \times C~')$ 

2) $comp2 = (f~' \times g~') \circ <f, g>:A \rightarrow (B~' \times C~')$.

\ph{exm}{lGammanotaKleislistrengthII} 
Let $MX = X + Exc$ be the exceptions monad of Example \ref{exceptionsarestrongmonads}, where 
$Exc$ now has exactly two elements $e1$ and $e2$. For arbitrary fixed values $a$ in $A$ and $b$ in $B$, let $f(a) = \eta(b)$ (where $\eta$ is the return method), $g(a) = e2$, $f~'(b) = e1$ and $g~'$ is arbitrary, all with the same type structure as above but where the maps are now in $\kl{C}{M}$, so for instance $f:A \rightarrow B$ in $\kl{C}{M}$ corresponds to $f:A \rightarrow MB$ in \textbf{C}. Interpreting the compositions in the Kleisli category $\kl{C}{M}$, $comp1:A \rightarrow M(B~' \times C~')$ and $comp2:A \rightarrow M(B~' \times C~')$ should agree. Using the derived left prestrength $lst\Gamma$ on value $a$, $comp1(a) = e2$ while $comp2(a) = e1$ and so the compositions disagree. In a symmetric fashion one can show that the derived right prestrength also fails to guarantee equality. The reader will recall that the derived right prestrength construction is what is usually referred to as the monad strength in the sense found in \cite{articleL} and thus a product semantics with this notion of strength will fail to be correct. In contrast either Kleisli strength in 
Example \ref{exceptionsarestrongmonads} will guarantee equality of the two compositions. 
\end{exm}

The above discussion illustrates that the prestrength construction leading to the usual notion of monad strength does not generally support products in the Kleisli category, not just for the exception monad but for many other monads such as the list, state, continuation and selection monads. In contrast Kleisli strengths exist for many of these monads. 

\subsection{Classical Commutative Monads and Kleisli Strength}
In the next several sections we investigate the connections between the previous results and commutative monads as well as monad compositions, where supporting
results can be found in \cite{articleD}. To emphasize the connection to prior work we briefly work with symmetric monoidal categories.

\ph{lem}{getGammafromlstrst}
If $\Gamma$ is an arbitrary Kleisli strength of order 2 on monad $\mathbf{K} = (K, \nu, \rho)$, define natural transformations $l\Gamma$, $r\Gamma$ by 
\begin{eqnarray*}
l\Gamma_{AB} &=& KA \otimes B \csr{1 \otimes \rho_B} KA \otimes KB \csr{\Gamma_{AB}} K(A \otimes B)\\
r\Gamma_{AB} &=& A \otimes KB \csr{\rho_A \otimes 1} KA \otimes KB \csr{\Gamma_{AB}} K(A \otimes B)
\end{eqnarray*}
Then 
\begin{eqnarray}
\Gamma_{AB} &=& \label{Gammafromlstrst} KA \otimes KB \bsr{l\Gamma_{A,KB}} K(A \otimes KB) \bsr{\monadext{(r\Gamma_{AB})}} K(A \otimes B)
\end{eqnarray}

Similarly $\Gamma_{AB} = (l\Gamma_{AB})^{\#}\circ r\Gamma_{KA,B}$
\end{lem}

Note that the lemma begins with an arbitrary Kleisli strength $\Gamma$ of order 2 on monad $\mathbf{K}$ which induces via the unit, $\rho$, the maps $l\Gamma$ and $r\Gamma$. In essence what the lemma states is that if we use the prestrength construction of section 3.3, the resulting prestrengths agree with each other and generate the original Kleisli strength. Thus the notion of Kleisli strength can be seen as a natural generalization of Kock's commutative monads to symmetric monoidal categories that are not necessarily closed and where the monad is no longer assumed to be a monoidal functor. This is important as very few monads of interest to computer science are monoidal functors and so we work instead with Kleisli strength. 

Checking whether a given monad has a Kleisli strength however, namely that conditions ($\Gamma A$) and ($\Gamma B$) hold, can often be tedious and complicated, so it would be useful to have an easier characterization for determining when Kleisli strength exists. In general this is not known but in the special case of the prestrength construction we can generalize a result of Kock. First a useful lemma.

\ph{lem}{rstisaKleisliLift} Given $st$-monad $(\textbf{C},M,st)$, the corresponding maps $rst$ and $lst$ form Kleisli lifting transformations.
\end{lem}
The proof is a straightforward diagram chase and simply checks that the following two equations hold (in the case of $rst$).

1) $rst \circ (1 \times \eta) = \eta$

2) $\mu \circ M(rst) \circ rst = rst \circ (1 \times \mu)$

\ph{thm}{KleisliequalsCommutative}
Given $st$-monad $(\textbf{C},M,st)$, then the derived prestrength $lst\Gamma$ is a Kleisli strength if and only if $\mathbf{M}$  is commutative ($lst\Gamma = rst\Gamma$) in which case $rst\Gamma$ is also a Kleisli strength. Likewise the symmetric statement holds for $rst\Gamma$.
\end{thm}

\begin{proof}
We look at the case of $lst\Gamma_{AB}= rst^{\#} \circ lst: MA \times MB \rightarrow M(A \times B)$. 
Condition ($\Gamma A)$ holds since

$lst\Gamma (\eta_A \times \eta_B)$ $= rst^{\#} \circ lst \circ (\eta \times 1) \circ (1 \times \eta)$

$= (rst^{\#} \circ \eta \circ (1 \times \eta)$ (by \ref {rstisaKleisliLift})

$= \mu \circ \eta \circ rst \ (1 \times \eta)$ (by naturality of rst)

$= \eta_{A \times B}$ (by \ref {rstisaKleisliLift}).

\medskip

Condition ($\Gamma B)$ holds since

$lst\Gamma (\mu \times \mu)$ $= rst^{\#} \circ lst \circ (1 \times \mu) \circ (\mu \times 1)$ (by definition)

$= rst^{\#} \circ M(1 \times \mu)\circ lst \circ (\mu \times 1)$ (naturality of lst)

$= \mu \circ \mu \circ MM(rst) \circ M(rst)\circ lst \circ (\mu \times 1)$
(by \ref {rstisaKleisliLift})

$= \mu \circ \mu \circ MM(rst) \circ M(rst) \circ \mu \circ M(lst) \circ lst$
(by \ref {rstisaKleisliLift})

$= \mu \circ \mu \circ MM(rst) \circ \mu \circ MM(rst) \circ M(lst) \circ lst$
(by naturality of $\mu$)

$= \mu \circ \mu \circ MM(rst) \circ \mu \circ MM(rst) \circ M(lst) \circ lst$
(by naturality of $\mu$)

$= \mu \circ M(rst) \circ \mu \circ \mu \circ MM(rst) \circ M(lst) \circ lst$
(by naturality of $\mu$)

$= \mu \circ M(rst) \circ \mu \circ M(\mu) \circ MM(rst) \circ M(lst) \circ lst$
(by property of monad)

$= \mu \circ M(rst) \circ \mu \circ M(\mu) \circ MM(lst) \circ M(rst) \circ lst$
(since $lst\Gamma = rst\Gamma$)

$= \mu \circ M(rst) \circ \mu \circ \mu \circ MM(lst) \circ M(rst) \circ lst$
(by property of monad)

$= \mu \circ M(rst) \circ \mu \circ M(lst) \circ \mu  \circ M(rst) \circ lst$
(by naturality of $\mu$)

$= \mu \circ \mu \circ MM(rst) \circ M(lst) \circ \mu  \circ M(rst) \circ lst$
(by naturality of $\mu$)

$= \mu \circ M(\mu) \circ MM(rst) \circ M(lst) \circ \mu  \circ M(rst) \circ lst$
(by property of monads)

$= \mu \circ M(lst\Gamma) \circ lst\Gamma$

\medskip
Conversely, suppose that the derived prestrength $lst\Gamma$ is a Kleisli strength where $lst\Gamma = rst^{\#} \circ lst$. Consider the corresponding maps $lst^* = lst\Gamma \circ (1 \times \eta )$ and $rst^* = lst\Gamma \circ (\eta  \times 1)$. We have that $lst^* = lst\Gamma \circ (1 \times \eta) = rst^{\#} \circ M(1 \times \eta) \circ lst = \mu \circ M\eta \circ lst = lst$. Also $rst^* = lst\Gamma \circ (\eta  \times 1) = rst^{\#} \circ lst \circ (\eta  \times 1) = rst^{\#} \circ \eta = \mu \circ \eta \circ rst = rst$. Thus by the argument of Lemma  \ref{getGammafromlstrst}, the maps $rst^*$ and $lst^*$ arising from $lst\Gamma$ agree in turn with $rst$ and $lst$ respectively and so generate both $lst\Gamma$ and $rst\Gamma$ which agree. Thus the monad is commutative.
\end{proof}

\smallskip
Note: The last result does not characterize whether a given monad has a Kleisli strength. For example the state monad is not commutative but does have a Kleisli strength as we note in the next example, rather it asserts that the derived prestrength is a Kleisli strength exactly when the monad is commutative, often a much easier condition to check.

\medskip

\ph{exm}{statemonadsarenotKockstrong} Consider the state monad $M$ where $MA = (A \times S)^S$. Building the left and right pre-strength constructions $lst\Gamma_{AB},rst\Gamma_{AB}:MA \times MB \rightarrow M(A \times B)$ we have,

\noindent
$lst\Gamma_{AB}(t_A,t_b)= \lambda s.let~(a,s_A) = t_A(s),~let(b,s_{AB}) = t_B(s_A)~in~(a,b,s_{AB})$\\
\noindent
$rst\Gamma_{AB}(t_A,t_b)= \lambda s.let (b,s_B) = t_B(s), ~let
(a,s_{BA}) = t_A(s_B) ~in~ (a,b,s_{BA})$. 

\noindent By the previous theorem since $lst\Gamma$ and $rst\Gamma$ don't agree, neither is a Kleisli strength, nor commutative. On the other hand $\Gamma_{AB}(t_A,t_b) = \lambda s.let (a,s_A) = t_A(s),~ let(b,s_{B}) = t_B(s)~ in (a,b,s)$ is a Kleisli strength. 
\end{exm}

\ph{exm}{exceptionsarenotKockstrongmonads} In Example \ref{exceptionsarestrongmonads} it was shown the exceptions monad $MA = X + Exc$ came equipped with different Kleisli strengths for non-trivial $Exc$. $M$ is not commutative however since for $a,b \in Exc$, the derived prestrengths are $lst\Gamma(a,b) = a$ while $rst\Gamma(a,b) = b$, and so $lst\Gamma \neq rst\Gamma$. 
\end{exm}

\subsection{Composing Monadic Data Types}

As the previous sections illustrate, monads play an important role in different aspects of functional programming. Since we often wish to apply more than one monad to a particular construction, finding circumstances where they compose is of particular interest. Monads compose exactly in the presence of a distributive law which is defined next. Finding such laws can be difficult in general, but in the presence of Kleisli strength a large class of these laws exists.

\ph{dfn} {DefnofDistLaw} For monads $\mathbf{H} = (H, \mu, \eta)$ and $\mathbf{K} = (K, \nu, \rho)$ on category \textbf{C}, a \textbf{distributive law of H over K} is a natural transformation $\lambda : HK \rightarrow KH$ for which the following diagrams commute.  
\begin{center}
\xext=3500 \yext=1700
\adjust[`\mu;`n;`Tn;`\mu]
\begin{picture}(\xext,\yext)(\xoff,\yoff)
\putmorphism(0,1600)(1,0)[H`HK` H\rho ]{975}1a
\putmorphism(1150,1600)(1,0)[`HKK` H\nu ]{850}{-1}a
\putmorphism(1050,930)(1,0)[`KKH` \nu H]{975}{-1}b
\putmorphism(975,1600)(0,-1)[`KH`\lambda]{660}1r
\putmorphism(2015,1600)(0,-1)[`` \lambda K]{330}1r
\putmorphism(2015,1275)(0,-1)[KHK``K\lambda ]{330}1r
\putmorphism(0,1600)(4,-3)[``]{980}1a
\put(450,1150){\makebox(0,0){$ \rho H$}}
\put(650,1350){\makebox(0,0){$(DL~A)$}}
\put(1500,1350){\makebox(0,0){$(DL~B)$}}
\put(2900,1350){\makebox(0,0)}
\putmorphism(0,600)(1,0)[K`HK` \eta K]{975}1a
\putmorphism(1150,600)(1,0)[`HHK` \mu K]{850}{-1}a
\putmorphism(1050,-70)(1,0)[`KHH`K \mu  ]{975}{-1}b
\putmorphism(975,600)(0,-1)[`KH`\lambda]{660}1r
\putmorphism(2015,600)(0,-1)[``H\lambda  ]{330}1r
\putmorphism(2015,275)(0,-1)[HKH`` \lambda H]{330}1r
\putmorphism(0,600)(4,-3)[``]{980}1a
\put(450,150){\makebox(0,0){$K\eta $}}
\put(650,350){\makebox(0,0){$(DL~C)$}}
\put(1500,350){\makebox(0,0){$(DL~D)$}}
\put(2900,350){\makebox(0,0)}
\end{picture}
\end{center} 
\end{dfn}
\vspace{.2in}

We now return to the problem of providing a means of composing monadic types such as $m[a]$. The key point is the existence of a distributive law. As pointed out in \cite{articleC}, a distributive law of $\mathbf{H}$ over $\mathbf{K}$ produces the composite monad  $(KH, (\nu \mu)(K \lambda H), \rho \eta)$ where the four laws above ensure the laws for a monad hold, thus providing a means of composing monadic data types. In particular, there may be more than one such law $\lambda$ for given monads $\mathbf{H}$ and $\mathbf{K}$ and it is critical to provide $\lambda$ when asserting the composition. In particular, monads equipped with Kleisli strength play an important role in the composition with free monadic data types. We address this next.

\ph{thm}{DLKleislistrengthoverfreemonad}
Let $\Sigma$ be a monadic finitary signature in $\mathcal{V}$ with corresponding free monad 
$(\Sigma^@,\mu,\eta)$. If $\mathbf{K} = (K,\nu,\rho)$ is a monad in $\mathcal{V}$ equipped with a family of Kleisli strengths $(\Gamma^\omega : \omega \in \Sigma_n)$ with each $\Gamma^\omega$ of order $n$ if $\omega \in \Sigma_n$, then there exists a distributive law 
$\lambda : \Sigma^@ K \rightarrow K \Sigma^@$.
\end{thm}

The categorical proof of this result is lengthy and will be skipped, however in section 4.1 we will sketch a proof in Haskell. This last result, and corresponding ones associated with linear monadic finitary signatures, generate compositions involving monadic data types such as lists, trees and exceptions which would not exist under normal circumstances, such as the list monad composed with itself, incorrectly claimed in \cite{articleA}. As emphasized earlier, Kleisli strength is a more general notion than commutativity and thus distributive laws can exist even for non-commutative monads. Since every commutative monad is Kleisli strong we have the immediate corollary.

\ph{cor}{DLCommutativeMonadoverfreemonad}
If in Theorem \ref{DLKleislistrengthoverfreemonad} $\mathbf{K}$ is commutative, there exists a distributive law $\lambda: \Sigma^@K \rightarrow K\Sigma^@$ of $\Sigma^@$ over \textbf{K}.
\end{cor}

\ph{exm}{powersetVcompose}
The monad $\mathbf{K} = \mathbf{P_0}$  of Example \ref{powersetKleislistrength} has a Kleisli strength of order n. By \ref{DLKleislistrengthoverfreemonad}, there exists a distributive law $\lambda : \Sigma^@ P_0 \rightarrow P_0 \Sigma^@$ for any monadic finitary signature $\Sigma$. For instance, if $\Sigma^@$ is the binary tree monad $V$, then  $\lambda:VP_0A \rightarrow P_0VA$ defined by  $\lambda E = \{E\}$, $\lambda(L(A_0)) = \{L(a) : a \in A_0\}$ and $\lambda (N(s,t)) = \{N(s_i,t_j): s_i \in \lambda(s), t_i \in \lambda(t)\}$ is a distributive law resulting in the composite monad $P_0V$. 
\end{exm}

\ph{exm}{readerVcompose} Similarly, for the reader monad $M$ of Example \ref{exponentialisstrong}, there exists a distributive law $\lambda : \Sigma^@ M \rightarrow M \Sigma^@$ for any monadic finitary signature $\Sigma$.  Now when $\Sigma^@$ is $V$, the distributive law $\lambda$ is defined to be $\lambda(E) = \lambda a.E$,  $\lambda(L(f)) = \lambda a.(L(f a))$ and $\lambda((N(t_1,t_2)) = \lambda a.(N((\lambda t_1)a, (\lambda t_2)a))$, resulting in the composite monad $MV$.
\end{exm}

The prior results provide a roadmap for constructing composite monads. They also provide a means for testing whether such compositions are correctly defined. The following example is a case in point. 

\ph{exm}{seqNotCorrect}
An important monad operation in Haskell is the function $sequence:: [m~a] \rightarrow m[a]$ i.e. a function with type structure $LMA \rightarrow MLA$. Since monads are at work, it is reasonable to require that the interpretation of this operation in the Kleisli category (for monad $M$) is consistent. As a functor, $L$ acting on function $f:A \rightarrow B$ is just the ordinary $map$ function, i.e. $map:: (a \rightarrow b)\rightarrow ([a] \rightarrow [b])$ in Haskell. If we now take into account arbitrary monad $M$, does $map$ preserve the composition of functions in $\kl{C}{M}$, namely can $L$ (or $map$) now be lifted to $\kl{C}{M}$ where the operation $sequence$ comes into play? If we denote the lift of $map$ (or $L$ on the function $f:a \rightarrow m~b$ as $\overline{f}:[a] \rightarrow m[b]$, then $\overline{f}$ is defined as $sequence.(map~f)$ and preservation of Kleisli composition requires that the property $\overline{g.f} = \overline{g}.\overline{f}$ should hold, which it does not. We supply two simple examples in Haskell next.
\end{exm}

\ph{exm}{SequenceNotSemanticallyCorrectI} Let $M$ be the list monad where $f~n = [n, 2 * n]$, and $g~n = [n, 3 * n]$, then the compositions 
$\overline{g.f}([1,3]) = [[1,3],[1,9],[1,6],...]$ and  $\overline{g}.\overline{f}([1,3]) = [[1,3],[1,9],[3,3]..]$ disagree.
\end{exm}

\ph{exm}{SequenceNotSemanticallyCorrectII}  Let $M$ denote the {\it IO} monad in Haskell, $f: Int \rightarrow IO~Char$, $g: Char \rightarrow IO~()$ be defined where $f(n)$ is defined to call {\it getChar} $n$-times while $g$ is simply $putChar$. While both $\overline{g.f}$ and $\overline{g}.\overline{f}$ will return something of type $IO[()]$ for a given list of integers, the resulting actions will differ. For example, we have that $\overline{g.f}([3,1,2])$ on keyboard input $abc,~d,~ef$ results in $abccddeff$ while $\overline{g}.\overline{f}$ on the same input generates $abcdefcdf$ and thus the compositions do not agree.
\end{exm}

What went wrong in these last examples? The operation $sequence$ fails to be a distributive law because of the choice of monads $M$. This also is suggestive of why past examples of the list monad composed with itself were not correct. If the monad $M$ comes equipped with a Kleisli strength, however, than a distributive law exists and a consistent notion of composition is possible.

\section{Applying results in Haskell} As the prior section illustrates, satisfying the laws for Kleisli strength or commutativity is critical to correctly composing monadic data types via distributive laws. Proving that these laws hold, however, can at times be either difficult or nonintuitive, the proof of Theorem \ref{KleisliequalsCommutative} being a case in point. It would be very useful if these laws could be expressed in a more accessible form, such as a functional programming format, which might allow for easier testing and verification.  Additionally, we would like to generate concrete code that might generically build the default methods required for monad composition. In this section we provide several results in this direction using Haskell.

\subsection{Kleisli Strength and Distributive Laws in Haskell}

We begin by recalling some Haskell notation. Given a monad on \textbf{C}, $\mathbf{H} = (H, \mu, \eta)$, we note that $\eta$ correspond to the {\it return} method in Haskell while $\mu: HHA \rightarrow H$ corresponds to a {\it join} function defined in Haskell by $join ~ a ~ = ~ a >>= id$. The bind operation $>>=$ is a default method for the monad class in Haskell, but it can easily be seen to correspond categorically to $(a~>>=f)~ = join(fmap~f~a)$, or put in more categorical notation, $a~>>=f ~= ~\mu(Hf(a)) = ~f^{\#}a$ using Manes notation. A special case of $>>=$ is $>>$ defined by $m ~>> ~k = ~m >>= \lambda x \rightarrow k$ where x is not free in $k$.

As pointed out in \cite{ArticleK}, the do-syntax provides an intuitive and convenient way of writing an abbreviated form for applying multiple bind operations. The interpretation of do is captured in the following two rules:

\noindent $do ~ e1 ; ~ e2 = ~~ e1 ~>>~ e2$ \\
$do ~ p \leftarrow e1; ~ e2 = ~~ e1 ~>>= ~(\lambda p \rightarrow e2)$ \\

Given $st$-monad $(\textbf{C},M,st)$ recall that a monad is called commutative if the left and right prestrengths $lst\Gamma$ and $rst\Gamma$ generated by the pre-strength construction agree. We have immediately the following lemma.

\ph{lem}{DoExpressionForCommutative} $M$ is commutative exactly when the following two {\it do expressions} agree for any pair $(a,b)$ in $MA \times MB$.

$ do  \indent x \leftarrow a \indent =   \indent do \indent y \leftarrow b \\
\indent \indent \indent y \leftarrow b \indent \indent \indent \indent  x \leftarrow a \\
\indent \indent \indent return (x,y)\indent \indent \indent return (x,y)\\
$
\begin{proof}
The reader can easily check that the left (right) do-expresssions correspond exactly to the pre-strength constructions $lst\Gamma(a,b)$, ($rst\Gamma(a,b))$  for the given monad $M$.
\end{proof}
\end{lem}

\bigskip
We now consider an alternative proof in Haskell of Theorem \ref{KleisliequalsCommutative} of the prior section. Given monad $(\textbf{C},M,st)$ as in the theorem, then the derived prestrength $lst\Gamma$ is a Kleisli strength if conditions ($\Gamma A)$ and ($\Gamma B)$ of Kleisli strength hold. Focusing on ($\Gamma B)$ we have the following result.

\ph{lem}{DoExpressionForKleisli} Applying the clockwise composition of condition ($\Gamma B)$ on value $(a,b)$ in $MMA \times MMB$ results in the following do expression in Haskell.\\
$ do \indent x \leftarrow a \\
\indent \indent y \leftarrow b \\
\indent \indent u \leftarrow x \\
\indent \indent v \leftarrow y \\
\indent \indent return (u,v)$\\
\begin{proof}

\end{proof}
\end{lem}

Going clockwise for $(a,b)$ and using Lemma \ref{DoExpressionForCommutative} we have
$\mu \circ M(lst\Gamma) \circ lst\Gamma (a, b)$ is exactly $\mu \circ M(lst\Gamma)$ applied to \\ 
$ do \indent x \leftarrow a \\
\indent \indent y \leftarrow b \\
\indent \indent return (x,y)$\\

Since $lst\Gamma$ is natural, $M(lst\Gamma)\circ return (x,y) = return \circ lst\Gamma (x,y)$ and again using Lemma \ref{DoExpressionForCommutative} results in $\mu$ applied to \\
\medskip

$  do \indent x \leftarrow a \\
\indent \indent y \leftarrow b \\
\indent \indent return (x,y) \\
\indent \indent do \indent u \leftarrow x \\
\indent \indent \indent \indent v \leftarrow y \\
\indent \indent \indent \indent return (u,v)$

By monad laws, applying $\mu$ simply removes the middle return.\\

\ph{lem}{DoExpressionForKleisli2} The counterclockwise composition of condition ($\Gamma B)$ on value $(a,b)$ in $MMA \times MMB$ results in the following do expression in Haskell.\\

$do \indent x \leftarrow a \\
\indent \indent u \leftarrow x \\
\indent \indent y \leftarrow b \\
\indent \indent v \leftarrow y \\
\indent \indent return (u,v)$\\

\begin{proof}
The counterclockwise composition applied to $(a,b)$ is just $lst\Gamma \circ (\mu \times \mu)(a, b)$. The expression $\mu(a)$ corresponds to the do expression\\
$ do \indent x \leftarrow a \\
\indent \indent u \leftarrow x \\
\indent \indent return ~u $ \\
with a similar do expression for $\mu(b)$.\\
Applying Lemma \ref{DoExpressionForCommutative} gives the desired result where we drop the redundant do's.
\end{proof}

\end{lem}

We now apply the previous results to prove Theorem \ref{KleisliequalsCommutative}.

\ph{thm}{NewProofKleisliequalsCommutative} Given monad $(\textbf{C},M,st)$, then the derived prestrength $lst\Gamma$ is a Kleisli strength if and only if $\mathbf{M}$  is commutative
\end{thm}
\begin{proof} By Lemma \ref{DoExpressionForKleisli2} the counterclockwise composition of condition ($\Gamma B)$ on value $(a,b)$ in $MMA \times MMB$ corresponds to\\

$do \indent x \leftarrow a \\
\indent \indent u \leftarrow x \\
\indent \indent y \leftarrow b \\
\indent \indent v \leftarrow y \\
\indent \indent return (u,v)$\\

If $M$ is commutative, we can swap the middle two lines resulting in the do-expression found in Lemma \ref{DoExpressionForKleisli} and so condition ($\Gamma B)$ holds. Showing condition ($\Gamma A)$ holds is trivial and left to the reader. 

Conversely if $lst\Gamma$ is a Kleisli strength, then condition ($\Gamma B)$ holds. Applying the values 
\newline $(return~x, (M~return) ~y)$ to the clockwise composition generates a do expression equivalent to \\
$ do \indent v \leftarrow y \\
\indent \indent u \leftarrow x \\
\indent \indent return (u,v)$ \\
which must agree with applying the same values counterclockwise generating a do expression equivalent to \\
$ do \indent u \leftarrow  x \\
\indent \indent v \leftarrow y \\
\indent \indent return (u,v)$ \\
Since $x$ and $y$ were arbitrary values of type $M~A$ and $M~B$ respectively, $\mathbf{M}$  is commutative.
\end{proof}\\

We now turn our attention to the results on distributive laws found in section 3. We use Haskell to provide a more intuitive approach to Corollary \ref{DLCommutativeMonadoverfreemonad}. First we state a useful lemma that provides a definition in Haskell of the distributive laws generated by the corollary.

\ph{lem}{DefnOfLambdaUsingDo} Let $(\Sigma^@,\mu,\eta)$ be the free monad of Theorem \ref{DLKleislistrengthoverfreemonad} with commutative monad $(K,\nu,\rho,st)$ both defined on $\textbf{C}$. Then the corresponding distributive law 
$\lambda:\Sigma^@K \rightarrow K\Sigma^@$ can be defined as follows in Haskell. For each $\omega \in \Sigma_n$, $\lambda(\omega(x_1,...x_n)) =\\ 
do \indent a_1 \leftarrow \lambda ~x_1 \\
\indent \indent ...\\
\indent \indent a_n \leftarrow \lambda ~x_n \\
\indent \indent return ~(\omega(a_1,...a_n))\\
$
\begin{proof}
The interested reader can check the original proof found in \cite{articleD} for details of the construction of $\lambda$ which is defined recursively. The do expression is a straightforward translation of that construction.
\end{proof}
\end{lem}

\ph{thm}{NewProofCommutativeImpliesComposition}
If in Theorem \ref{DLKleislistrengthoverfreemonad} $(\textbf{C},K,\rho,\nu,st)$ is commutative, there exists a distributive law of $\Sigma^@$ over \textbf{K}, $\lambda:\Sigma^@K \rightarrow K\Sigma^@$ .
\end{thm}
\begin{proof}
For ease of discussion, we use a simple example of the free polynomial monad $(\Sigma^@,\mu, \eta)$, namely $V$ the non-empty binary tree monad of Example \ref{binarytreemonad}(we drop the $^+$) where a trivial tree (i.e. a leaf) with value $x$ is denoted by $L(x)$ and a tree consisting of left and right subtrees $v1$ and $v2$ by $N(v1,v2)$. We have four laws to check that a distributive law exists for $KV$.

Focusing on law $(DL~B)$ and starting with a term $N( ~tt1, ~tt2)$ of type $VKK$, going clockwise  
$(\nu_V \circ K\lambda \circ \lambda)(N( ~tt1, ~tt2))$ corresponds by Lemma \ref{DefnOfLambdaUsingDo} to $\nu_V \circ K\lambda$ applied to the do expression\\
$do \indent t1 \leftarrow \lambda ~tt1 \\
\indent \indent t2 \leftarrow \lambda ~tt2 \\
\indent \indent return ~(N( ~t1, ~t2))$\\ 

\noindent
Since return is natural, $(K \lambda \circ return) ~N( ~t1, ~t2) =~ return ( \lambda (N( ~t1, ~t2))) =~return$ applied to\\
$do \indent a \leftarrow \lambda ~t1 \\
\indent \indent b \leftarrow \lambda ~t2 \\
\indent \indent return ~N( ~a, ~b)$\\
composing with $\nu$ removes the middle $return$ and the redundant $do$, resulting in the do expression \\
$do \indent t1 \leftarrow \lambda ~tt1 \\
\indent \indent t2 \leftarrow \lambda ~tt2 \\
\indent \indent a \leftarrow \lambda ~t1 \\
\indent \indent b \leftarrow \lambda ~t2 \\
\indent \indent return ~N( ~a, ~b)$\\

In a similar fashion counterclockwise composition $\lambda \circ V(\nu)$ generates a do expression equivalent to \\
$do \indent t1 \leftarrow \lambda ~tt1 \\
\indent \indent a \leftarrow \lambda ~t1 \\
\indent \indent t2 \leftarrow \lambda ~tt2 \\
\indent \indent b \leftarrow \lambda ~t2 \\
\indent \indent return ~N( ~a, ~b)$\\

When $K$ is commutative, the two middle components in the latter do expression can be swapped resulting in the clockwise composition and so $(DL~B)$ holds. The two conditions $(DL~A)$ and $(DL~C)$ are easily verified leaving only $(DL~D)$. Since $V$ is a recursively defined data type, $\nu~N(tt1,tt2)= N(\nu~tt1,\nu~tt2)$. Since $\lambda$ is also defined recursively, condition $(DL~D)$ follows immediately by structural recursion. The final case of considering elements of the form $(L~ x)$ easily holds for all conditions and is left to the reader.
\end{proof}

\subsection{Some examples in Haskell}

\ph{exm}{ReturnToSequence}
One can generalize the last result to include quotients of free monads, such as the list monad. In this case the Haskell code for the distributive law $\lambda:[m~a] \rightarrow m[a]$ produced whenever $M$ is a commutative monad is defined by Lemma \ref{DefnOfLambdaUsingDo} as\\
$\lambda [~] = return [~]\\
\lambda (x:xs) = ~ do\\
\indent \indent \indent \indent \indent a \leftarrow x \\
\indent \indent \indent \indent \indent b \leftarrow \lambda ~xs \\
\indent \indent \indent \indent \indent return ~(a:b)$\\
which coincides exactly with the usual definition of $sequence$ found in Haskell. 
The problems arising in Examples \ref{SequenceNotSemanticallyCorrectI} and \ref{SequenceNotSemanticallyCorrectII} occurred since the monads used were the list and IO monad respectively, neither of which is commutative.
\end{exm}

\ph{exm}{GeneralizingSequenceToBinTrees} 
It is an easy matter to define a correct distributive law for composing binary trees over other commutative monads such as the reader monad, $\lambda:V(r \rightarrow a) \rightarrow (r \rightarrow V a)$. 
\end{exm}

Theorem \ref{DLCommutativeMonadoverfreemonad} indicates that one can build composite monadic data types for any free polynomial monad data types over commutative monads such as powerset, bag, reader, writer(for commutative monoid) and maybe monads. Lemma \ref{DefnOfLambdaUsingDo} provides a roadmap for defining the corresponding distributive laws in Haskell. What remains is to provide an explicit definition of the $bind$ and $return$ operators in Haskell so we can apply these results. We do this next.

\ph{pro}{DefiningBind} Given a recursive polynomial monad $V$ over a commutative monad $M$ one can define the resulting composite monad $MV$ generated by distributive law $\lambda$ in Haskell where the bind operation is defined by $(a >>= f)~ =  \\	
\indent  do \\
\indent  x \leftarrow  a \\
\indent  b \leftarrow  \lambda (fmap ~ f ~ x) \\
\indent  return (\indent do \\
\indent \indent \indent \indent \indent y \leftarrow  b \\
\indent \indent \indent \indent \indent 	v \leftarrow  y \\
\indent \indent \indent \indent \indent 	return  ~v \\
\indent \indent ) \\
$
\end{pro}
\begin{proof}
The return method ($\eta$) for $MV$ is trivial and simply the composition of the individual return methods, i.e. $return_M \circ return_V$. Exploiting basic propeties of a monad we arrive at a definition of bind on the composite monad $MV$. The join operation $join: MVMV \rightarrow MV$ is defined by $\mu \circ MM(\nu) \circ (M\lambda V)$ which in turn is used to define the bind operation on $MV$ via $(a >>= f) = join \circ (fmap~f ~a)$ where $fmap$ denotes the action of the functor on the new composite monad. Now a straightforward diagram chase shows that\\
$join \circ (fmap~ f) \circ \eta  =\\
\mu \circ MM(\nu) \circ (M\lambda_V) \circ (fmap ~ f) \circ \eta  =\\
\mu \circ MM(\nu) \circ (M\lambda_V) \circ \eta \circ (fmap ~f) ~ =\\
\mu \circ MM(\nu) \circ \eta \circ \lambda _V \circ (fmap ~f) ~ =\\
\mu \circ \eta \circ M(\nu) \circ \lambda_V \circ (fmap ~f) ~ =\\
M(\nu) \circ \lambda_V \circ (fmap ~f) ~$\\
where the latter uses of $fmap$ refer to the monad $V$.\\

\noindent So  $(a >>= f) ~$ corresponds to $M(\nu)$ applied to the do expression\\
$
\indent  do \\
\indent  x \leftarrow a\\
\indent  b \leftarrow  \lambda_V (fmap ~f ~x)\\
\indent  return (\nu ~ b)$ \\

\noindent 
rewriting $\nu$ as a do expression and using naturality generates the desired result.

\vspace{.1in}
\end{proof}

\noindent
Next we provide several examples. A key point is that no actual work is required to decipher the details of how the operators $return$ and $bind$ behave for these examples since it is already determined generically by the prior results.

\ph{exm}{CompositeReaderBinaryTree} Exploiting the previous proposition, computing the bind method for the composite monad $Maybe[a]$ is straightforward. For instance\\
$Just [2,4,6])>>= f = Just [2,4,4,8,6,12]$ while\\
$Just [2,3,6]) >>= f = Nothing$ where\\ 
$f ~n = Just([n, 2*n])$ if $n$ is even and $Nothing$ otherwise.\\
\end{exm}

\ph{exm}{ComputingSequenceforBinTrees}
For the composite monad $r \rightarrow V a$ of the Example \ref{GeneralizingSequenceToBinTrees} we have that given \\ 
$f ~a = N( (t1 ~a), (t2 ~a))) \\
t1 ~a = N((L ~a), (L ~(a+1)))\\
t2 ~a = N ((L ~(2*a)), (L ~(3*a)))$\\
where $g ~n = \lambda a \rightarrow N ((L ~(2*a)) ,(L ~(4*a)))$ if n is even \\
\indent \indent \indent $ = \lambda a \rightarrow N( (L ~a), (L ~(3*a)))$ otherwise, \\
\noindent
$(f >>= g)(5) = N (N (N (L~ 5) (L ~15)) (N (L~ 10) (L ~20))) (N (N (L~ 10) (L~ 20)) (N (L ~5) (L ~15)))$.

\end{exm}

\ph{exm}{CompositeMSetBinaryTree} Consider the M-set monad $M X = C \times X$ of Example 
\ref{certificatemonadnotstrong} where C is a monoid. When $C$ is  commutative, so is $M$ and so we can form the composite monad $MVA = C \times VA$. Here the definition of $\lambda:V(C \times A) \rightarrow C \times VA$ derived from Lemma \ref{DefnOfLambdaUsingDo} yields for instance\\
$\lambda ~N(~L(c_1,a), ~L(c_2,b)) = (c_1*c_2, N(~(L ~a), ~(L ~b)))$.\\
\\
Likewise using Proposition \ref{DefiningBind}, $f ~a = (a,N( ~L(2*a), ~L(3*a)))$ produces\\
$(5, N(~(L~ 3), ~(L ~4))) >>= f = (12~, N~(~N(~(L ~6), ~(L ~9)), ~N(~(L ~8), ~(L~12)~)))$ \\
where $C$ is the commutative monoid of integers under addition.
\end{exm}

\section{Summary and Future Work}
The results of section 3 indicate that there are tradeoffs on how to utilize monads to interpret products semantically. Derived strengths are a natural consequence of the existence of an st-monad however they may fail to provide a correct semantics except in the exceptional case when the monad is commutative. Kleisli strengths provide an alternative approach. They generalize the notion of commutative and are thus applicable to a much wider array of monad used in computing while also guaranteeing that products lift to the coresponding Kleisli categories. An important application comes up in the context of monad composition where derived strength is not sufficient, but Kleisli strength is, to guarantee the proper composition of certain monads with polynomial monadic data types. Additional restrictions on the class structure of monads along the lines of commutativity or Kleisli strength will thus be required to ensure composition of functions in Kleisli. Section 4 provides an approach to earlier results using Haskell. In addition to providing explicit expressions that can test for the existence of Kleisli strength and distributivity, it also provides the generic code required for the default methods for a monad composition class for the compositions investigated.

Other constructs related to composing monads arise and might profit from the approach introduced in the paper. For instance, the previous section included an example related to the monad transformer construction. In \cite{articleH} the connection between the present work on composing monads and monad transformers is explored in more detail. Another important question is understanding when Kleisli strength is preserved under monad composition. While an initial result can be found in \cite{articleD}, in \cite{articleF} we provide additional examples and results that exploit the Haskell-centric approach found in this paper.

\section{Acknowlegement}

This paper is dedicated to David Schmidt on the occasion of his 60th birthday. David has been a friend and colleague for many years, someone who has shown consistent care and support for his students, colleagues, and the research community. I wish David continued success for many years to come.

\nocite{*}
\bibliographystyle{eptcs}
\bibliography{NotionsOfMonadStrength}
\end{document}